\documentclass[useAMS,usegraphicx,usenatbib]{mn2e}
\usepackage{rotating}
\def\degrees{^\circ}

\def\arcsec{^{\prime\prime}}
\def\kms{\mathrm{km\;s^{-1}}}

\def\kpc{{\rm\,kpc}}

\def\eg{{\it e.g.\ }}

\def\cf{{\it cf.\ }}

\def\spose#1{\hbox to 0pt{#1\hss}}

\newcommand{\lta}{\mathrel{\raise2pt\hbox{\rlap{\hbox{\lower6pt\hbox{$\sim$}}}\hbox{$<$}}}}
\newcommand{\gta}{\mathrel{\raise2pt\hbox{\rlap{\hbox{\lower6pt\hbox{$\sim$}}}\hbox{$>$}}}}

\def\ud{{\mathrm d}}

\title[Dark matter content and internal dynamics of NGC 4697]{Dark matter content and internal
  dynamics of NGC 4697: NMAGIC particle models from slit data and planetary nebulae velocities} 

\author[F. De Lorenzi et al.]{Flavio De
  Lorenzi$^{1,2}$\thanks{E-mail:lorenzi@mpe.mpg.de}, Ortwin
  Gerhard$^1$, Roberto P. Saglia$^1$, Niranjan Sambhus$^{2}$ \and
  Victor P. Debattista$^{3}$\thanks{RCUK Academic Fellow}, Maurilio
  Pannella$^1$, Roberto H.  M{\'e}ndez$^4$ \\ $^1$ Max-Planck-Institut
  f\"ur Ex. Physik, Giessenbachstra{\ss}e, D-85741 Garching, Germany
  \\ $^2$ Astron.  Institut, Universit\"at Basel, Venusstrasse 7,
  Binningen, CH-4102, Switzerland \\ $^3$ Centre for Astrophysics,
  University of Central Lancashire, Preston, PR1 2HE, UK \\ $^4$
  Institute for Astronomy, University of Hawaii, 2680 Woodlawn Drive,
  Honolulu, HI 96822, USA}

\begin{document}   
   
\date{Accepted ---. Received ---; in original form ---}

\pagerange{\pageref{firstpage}--\pageref{lastpage}} \pubyear{----}
  
\maketitle

\label{firstpage}

\begin{abstract}   
  We present a dynamical study of NGC 4697, an almost edge-on,
  intermediate-luminosity, E4 elliptical galaxy, combining new surface
  brightness photometry, new as well as published long-slit absorption
  line kinematic data, and published planetary nebulae (PNe) velocity
  data.  The combined kinematic data set extends out to $\simeq 5'
  \simeq 4.5 R_{e}$ and allows us to probe the galaxy's outer halo.

  For the first time, we model such a dataset with the new and flexible
  $\chi^2$-made-to-measure particle code NMAGIC.  We extend NMAGIC to
  include seeing effects, introduce an efficient scheme to estimate the
  mass-to-light ratio, and incorporate a maximum likelihood technique
  to account for discrete velocity measurements. 

  For modelling the PNe kinematics we use line-of-sight velocities
  and velocity dispersions computed on two different spatial grids,
  and we also use the individual velocity measurements with the
  likelihood method, in order to make sure that our results are not
  biased by the way we treat the PNe measurements.

  We generate axisymmetric self-consistent models as well models
  including various dark matter halos.  These models fit all
  the mean velocity and velocity dispersion data with $\chi^2/N<1$,
  both in the case with only luminous matter and in potentials
  including quite massive halos. The likelihood analysis together with
  the velocity histograms suggest that models with low density halos
  such that the circular velocity $v_c\lta200\kms$ at $5R_e$ are not
  consistent with the data. A range of massive halos with
  $v_c\simeq250\kms$ at $5R_e$ fit the PN data best. To derive
  stronger results would require PN velocities at even larger radii.
   The best fitting models are slightly radially anisotropic;
  the anisotropy parameter $\beta=\simeq0.3$ at the center, increasing
  to $\beta\simeq0.5$ at radii $\gta 2R_{e}$.

\end{abstract}   


\section{Introduction}   
\label{sec:introduction}   

The presence of dark matter (DM) has long been inferred around spiral
galaxies from their flat rotation curves, and galaxies are now
generally believed to be surrounded by extended dark matter halos.
Indeed, in the current $\Lambda$-cold dark matter ($\Lambda$CDM)
cosmology, galaxies form within the potential wells of their halos.
The standard picture for the formation of elliptical galaxies is
through mergers of smaller units. Ellipticals should thus also be
surrounded by dark matter halos. Their halos are particularly
interesting because ellipticals are among the oldest galaxies and are
found in the densest environments.

Unfortunately, mass measurements in elliptical galaxies have been
difficult because of the lack of a suitable ubiquitous tracer such as
neutral hydrogen rotation curves in spirals.  In giant ellipticals,
there is evidence for dark matter from X-ray emission
\citep[\eg][]{awaki+94,loewenstein+99,humphrey+06} and gravitational
lensing \citep[\eg][]{griffiths+96, treu+koop04, rusin+kochanek05}. In
more ordinary ellipticals, mass estimates come from stellar dynamical
studies, which have been limited by the faintness of the galaxies'
outer surface brightness to radii less than two effective radii from
the centre, $R \lta 2R_{e}$ \citep[\eg][]{krona_etal00, thomas+07}.
These studies suggest that the dark matter contributes $\sim10-40\%$
of the mass within $R_e$ \citep{gerhard+01, cappellari+06}, consistent
with the lensing results.

The strong emission line at $[\mathrm{OIII}]\lambda 5007$ from
planetary nebulae offers a promising tool to overcome this limitation
and to extend stellar kinematic studies to larger radii
\citep{hui_etal95,tremblay+95,arnaboldi+96,arnaboldi+98}.  Also, in
the less massive, X-ray faint ellipticals, PNe may be the primary tool
for constraining the dynamics at large radii.  Once the PNe are
identified, their line-of-sight velocities can be obtained from the
Doppler shift of the narrow emission line.  Interestingly, the derived
PNe dispersion profiles in the elliptical galaxies NGC 4697
\citep{mendez_etal01} and NGC 821, 3379 and 4494
\citep{romanowsky_etal03,douglas+07} were found to decline
significantly with radius outside $1 R_e$. Their spherically symmetric
dynamical analysis led \citet{romanowsky_etal03} to the conclusion
that these galaxies lack massive dark matter halos; however,
\citet{dekel+05} argued that the well known mass-anisotropy degeneracy
allows for declining dispersion profiles even when a standard DM halo
is present.

In the present paper we focus on NGC 4697, a normal and almost edge-on
E4 galaxy located along the Virgo southern extension.
\citet{mendez_etal01} obtained a planetary nebula luminosity function
(PNLF) distance of $10.5\pm 1\;\mathrm{Mpc}$ from magnitudes of $531$
PNe, and \citet{tonry_etal01} measured a surface brightness
fluctuation (SBF) distance of $11.7\pm 0.1\;\mathrm{Mpc}$.  This
fairly isolated galaxy has a total B magnitude $B_T=10.14$ and harbors
a central super massive black hole (SMBH) of mass $1.2\times 10^8
M_{\odot}$ \citep{pinkney_etal00}. A Sersic law with $R_{e} = 66$
arcsec gives a good fit to the surface brightness profile out to about
4 arcmin (see Section 2). Based on the disky isophote shapes
\citet{carter87} and \citet{goudfrooij_etal94} inferred a stellar disk
along the major axis. The contribution of the disk kinematics to the
major axis line-of-sight velocity distributions was estimated by
\citet{scorza+98}.
X-ray observations with ROSAT \citep{sansom_etal00} show a lack of
large scale hot gas in the halo of this galaxy. Using more recent
Chandra data, \citet{irwin_etal00} could resolve most of this emission
into non-uniformly distributed low mass X-ray binary (LMXB) point
sources, suggesting that NGC 4697 is mostly devoid of interstellar gas
and perhaps does not have substantial amounts of DM. 

Dynamical models of NGC 4697 have been constructed by
\citet{binney_etal90} and \citet{dejonghe_etal96}, both based on
photometry and absorption line kinematic data within $\sim 1 R_{e}$.
The data were consistent with a constant mass-to-light-ratio and none
of these models showed evidence for dark matter. \citet{mendez_etal01}
obtained velocities for $531$ PNe and derived a velocity dispersion
profile out to approximately $4.5 R_e$.  Assuming an isotropic
velocity distribution, \citet{mendez_etal01} found that the PNe
velocity dispersion profile is consistent with no DM inside $4.5
R_{e}$, but that DM can be present if the velocity distribution is
anisotropic. This was also argued by \citet{dekel+05} to be the main
cause of the finding by \citet{romanowsky_etal03}, that their three
intermediate luminosity galaxies lacked significant dark matter halos
\citep[but see][]{douglas+07}.  Contrary to these three galaxies,
which are nearly round on the sky, NGC 4697 is strongly flattened and
likely to be nearly edge-on, thus easier to model since shape
degeneracies are much less severe.

In the light of this, it is important enough to perform a further
careful analysis of this galaxy. In this paper we construct dynamical
models of NGC 4697 with the very flexible NMAGIC particle code, making
use of new and published slit kinematics and the \citet{mendez_etal01}
PN data.
The paper is organized as follows. In Section \ref{sec:obsdata} we
describe our new observational data and all other data that are used
for the dynamical models. In Section~\ref{sec:nmagic} we give a brief
explanation of the NMAGIC modeling technique. We extend the method to
include seeing effects, introduce an efficient scheme to estimate the
mass-to-light ratio, and show how discrete velocity measurements may
be taken into account. In Section \ref{sec:isorot} we construct
isotropic rotator models to test and calibrate the method, preparing
for the dynamical modeling of NGC 4697 which is then performed in
Section \ref{sec:models}. Finally, the paper closes with our
conclusions in Section \ref{sec:conclusions}.

%
\section{Observational Data}
\label{sec:obsdata}
In this section, we describe the observational data used in the
present study, including new CCD photometry and new long-slit
absorption line kinematics. We also describe here the procedure
employed for the deprojection of the photometric data.
In the following we adopt a distance $10.5 \;\mathrm{Mpc}$ to NGC 4697
\citep{mendez_etal01}.
%
\begin{figure}
\centering 
\includegraphics[width=0.92\hsize,angle=-90.0]{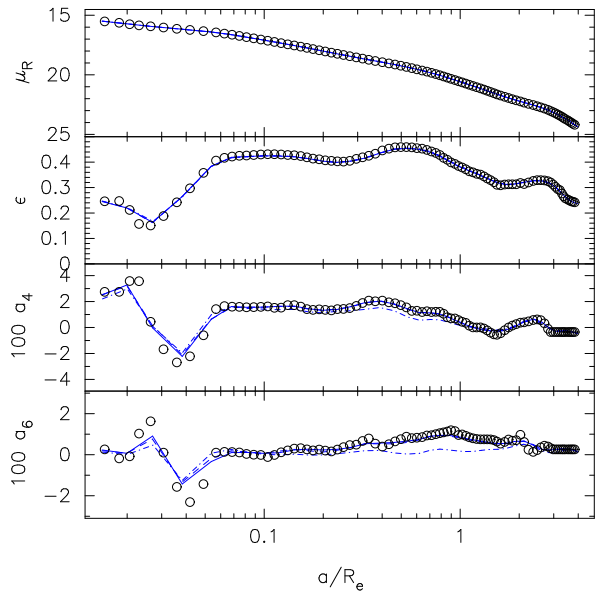}
\vskip0.5truecm
\caption[]{Comparison of the photometry of NGC 4697 with reprojected
  three-dimensional luminosity models. The data points correspond to
  the observed photometry for NGC 4697 (for the deprojection, the
  $a_4$ and $a_6$ values have been set to constants beyond $\simeq
  3R_e$). The solid line shows the edge-on deprojected model
  reprojected, the dashed line the $i=80^\circ$ deprojected model
  reprojected, and the dash-dotted line the $i=67^\circ$ model. The
  panels show, from top to bottom, surface brightness $\mu_R$,
  ellipticity $\epsilon$ and the isophotal shape parameters $a_4$ and
  $a_6$.}
\label{fig:photo}
\end{figure}
\subsection{Surface Photometry}
\label{sec:photo}
The R-band data used in the present work were taken in April 2000 as
part of the ESO Proposal 064.N-0030 (P.I. R.P. Saglia) at the Wide
Field Imager on the ESO-MPIA $2.2m$ telescope. Six 5 minutes, dithered
exposures where taken in sub-arcsec seeing conditions. After the usual
data reduction procedures (performed using the IRAF task mscred), the
data were tabulated as radial profiles of surface brightness $\mu$,
ellipticity $\epsilon$, position angle PA and Fourier shape
coefficients \citep{bender_moell87}. The surface brightness was
calibrated using the R band photometry of \citet{Peletier1990}. Systematic
errors due to sky subtraction ($\mu_{sky}=20.18$ Rmag arcsec$^{-2}$)
are always smaller than 10\%. Fig. \ref{fig:isophotes} shows the
derived profiles and Table \ref{tabphot4697} gives them in tabular
form.  The isophotes of NGC 4697 do not show any appreciable twist in
PA and have a positive $a_4$ coefficient in the galaxy's inner parts,
which is well-explained by a near-edge-on embedded disk
\citep{scorza+95}. The galaxy has some dust in the inner regions
\citep{pinkney_etal03}, but the R-band observations are relatively
unaffected by it.  The outer isophotes are progressively slightly off-centered.
A Sersic fit to the
surface brightness profile results in Sersic index $n=3.5$ and
effective radius $R_e=66$ arcsec. The older value of 95 arcsec from
\citet{binney_etal90} was based on photometry reaching only 120
arcsec; thus we use $R_e=66$ arcsec in the following. For a distance
of $10.5 \;\mathrm{Mpc}$ this corresponds to $3.36 \;\mathrm{kpc}$.

\subsection{Deprojection}
\label{sec:depro}
For our dynamical study, we will fit particle models to the
deprojected luminosity density using NMAGIC, \cf
Section~\ref{sec:nmagic}.  To obtain the luminosity density 
we need to deproject the surface brightness. This inversion
problem is unique only for spherical or edge-on axisymmetric systems
\citep{bin_tre87}. For axisymmetric systems inclined at an angle $i$,
the Fourier slice theorem \citep{rybicki87} shows that one can recover
information about the density only outside a ``cone of ignorance'' of
opening angle $90\degrees-i$. The deprojection of photometry for
galaxies with $i$ significantly less than $90\degrees$ can thus be
significantly in error because of undetermined konus densities
\citep{gerhard_bin96,kochanek_ryb96}. 

Fortunately, NGC 4697 is seen almost edge-on and hence does not suffer
from this ill-condition.  \citet{dejonghe_etal96} observed a nuclear
dust lane with a ring-like appearance, elongated along the major axis
of NGC 4697. Assuming that the ring is settled in the equatorial
plane, they estimated an inclination angle $i=78\degrees\pm
5\degrees$.  Applying a disk-bulge decomposition to the isophote
shapes of the galaxy, and assuming a thin disk, \citet{scorza+95}
derived an inclination $i=67\degrees$. This was updated by
\citet{scorza+98} to $i=70\degrees$.  \citet{scorza+95} also estimated
the velocity dispersion $\sigma_d$ of the disk component from the
major axis line-of-sight velocity distributions.  From their plots we
estimate $\sigma_d\simeq 95 \kms$ at the half-mass radius of the disk,
$r_D\simeq 13"$.  Assuming that the vertical dispersion in the disk is
comparable, we can estimate the intrinsic flattening of the disk
$\propto \sigma^2_d/v_c^2\simeq 0.2$, using the measured rotation
velocity. A disk with intrinsic thickness $h/R\simeq 0.2$ would give
the same isophote distortions for inclination $i\simeq 80\degrees$ 
as a thin disk with $i=67-70\degrees$, in agreement with the value
found by \citet{dejonghe_etal96}.

\begin{figure}
\centering 
\includegraphics[width=0.92\hsize,angle=-90.0]{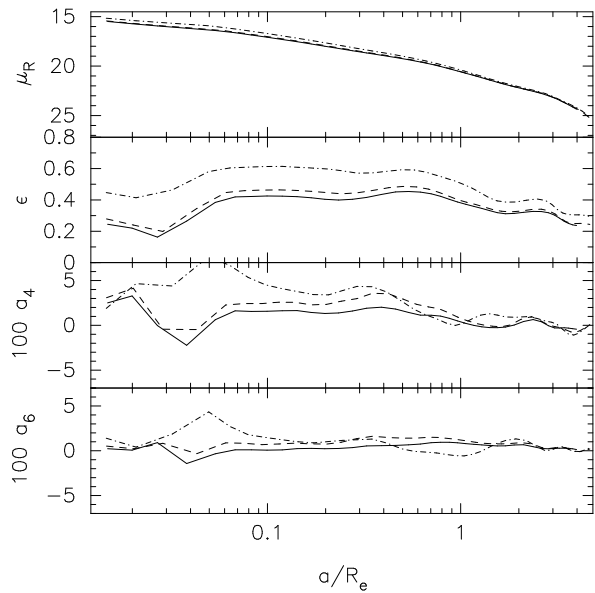}
\vskip0.5truecm
\caption[]{Isophote parameter profiles for the NGC 4697 photometry
  compared with those of the deprojected luminosity models shown in
  Figure \ref{fig:photo} but now as seen in edge-on projection. NGC
  4697 is coded as the solid line, whereas the $i=80^\circ$ and
  $i=67^\circ$ luminosity models are presented by the dashed and the
  dash-dotted lines, respectively. }
\label{fig:repro90}
\end{figure}

We have deprojected NGC 4697 for inclinations $i=90\degrees$,
$i=80\degrees$ and $i=67\degrees$, using the method and program of
\citet{magorrian99}. This algorithm uses a maximum penalized
likelihood method with a simulated annealing scheme to recover a
smooth three-dimensional luminosity density distribution which, when
projected onto the sky-plane, has minimal deviations from the observed
SB. The three-dimensional luminosity density so obtained extends
beyond the radial range of the data, where the penalized likelihood
scheme favours an outer power-law density profile.  Figure
\ref{fig:photo} compares the observed photometry with the three
deprojected luminosity models reprojected on the sky. In the range
from $0.2 R_e$ to $2 R_e$ the $i=67\degrees$ deprojection yields a
significantly less good fit to the observed $a_4$ and $a_6$.  Figure
\ref{fig:repro90} compares the radial run of the isophotal shape
parameters for the $i=80\degrees$ and $i=67\degrees$ luminosity models
projected edge-on, with the observed photometry of NGC 4697.  The
$i=80\degrees$ deprojection produces again somewhat better results. It
is also more physical because it allows for a finite thickness of the
inner disk of NGC 4697, as discussed above.  Hence we will adopt it
for the dynamical study of NGC 4697 to follow.

\subsection{Kinematic Data}
\label{sec:kindata}
\subsubsection{Stellar-absorption line data}

Long slit absorption line kinematics within $\sim R_{e}$ have been
reported, amongst other works, by \citet{binney_etal90} and
\cite{dejonghe_etal96}.  The \citet{binney_etal90} kinematic data (BDI
data) consist of line-of-sight velocity $v$ and velocity dispersion
$\sigma$ profiles along the major axis, along slits 
$10\arcsec$ and $20\arcsec$ parallel to the major axis, along the
minor axis, and along a slit $22\arcsec$ parallel to the minor axis.
They are derived using the Fourier Quotient (FQ) method \citep{IS82}.
\citet{dejonghe_etal96} have published further $v$ and $\sigma$
measurements (DDVZ data) at various position angles, also measured
with the FQ method.

\begin{figure*}
\includegraphics[width=0.49\hsize,angle=0.0]{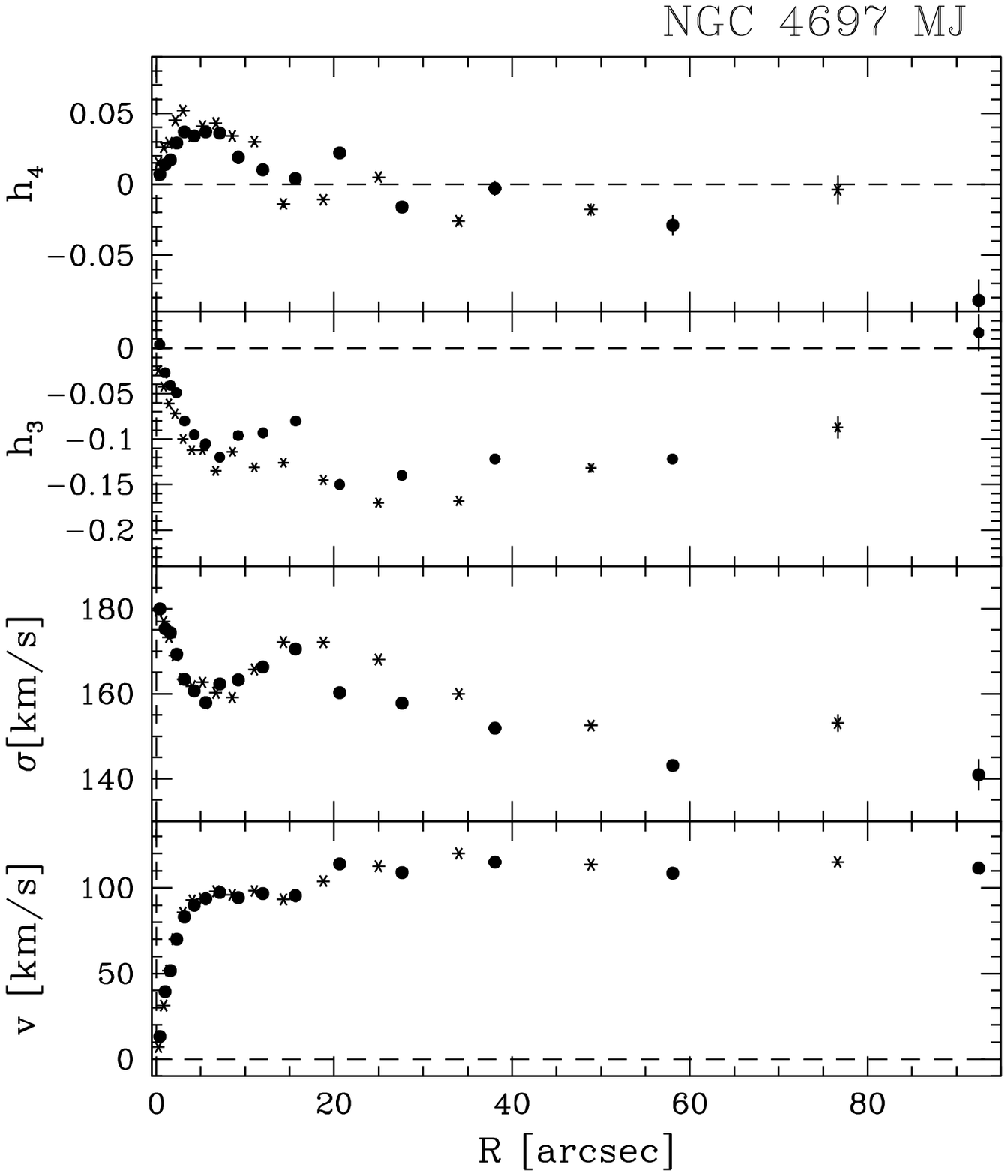}
\includegraphics[width=0.49\hsize,angle=0.0]{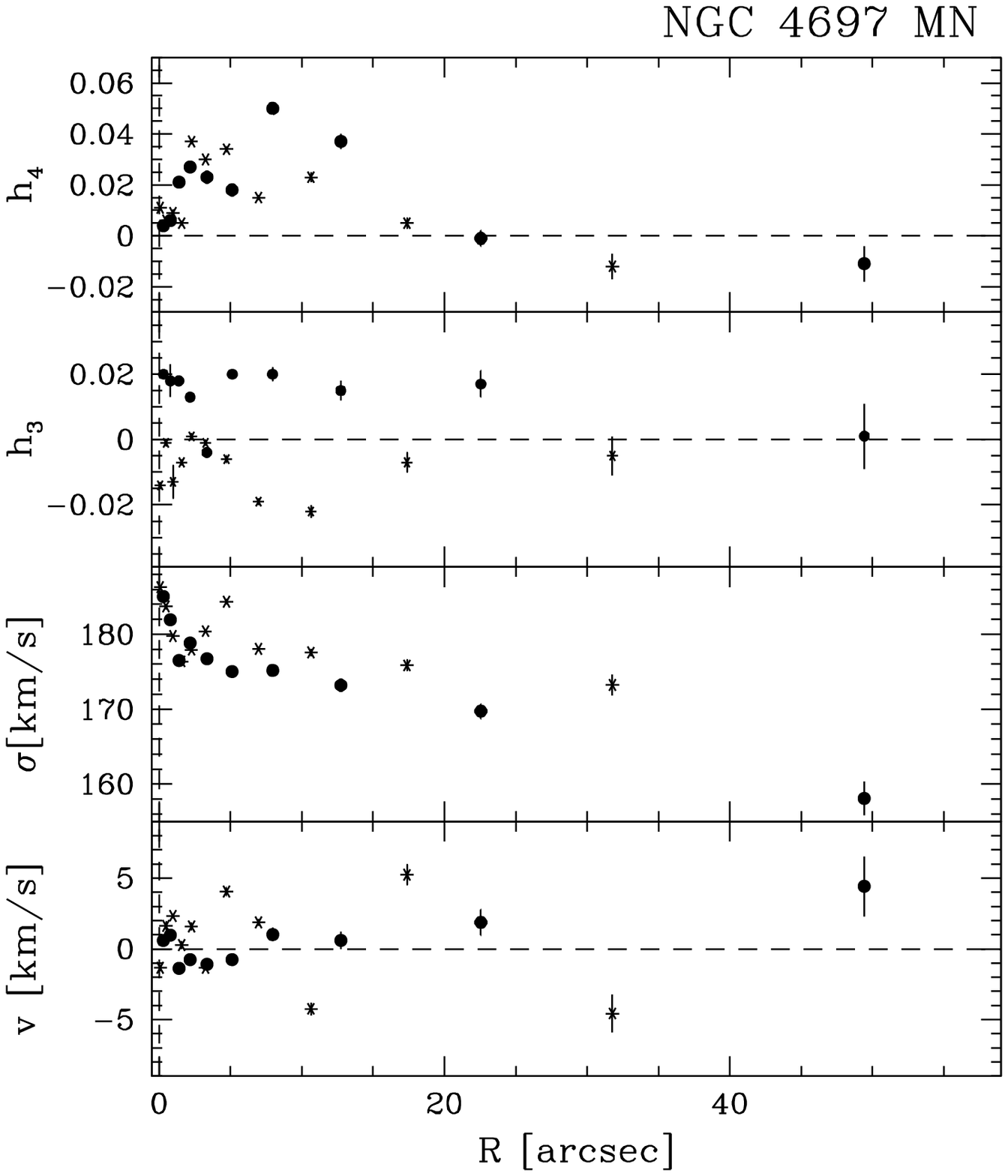}
\vskip0.5truecm
\caption[]{The kinematics along the major (left panel) and minor
  (right panel) axis of NGC 4697.  The filled and starred symbols
  refer to the data folded along the axes.}
\label{fig:kindata}
\end{figure*}

\begin{figure*}
\includegraphics[width=0.47\hsize,angle=0.0,clip=true,viewport=95 37 540 542]{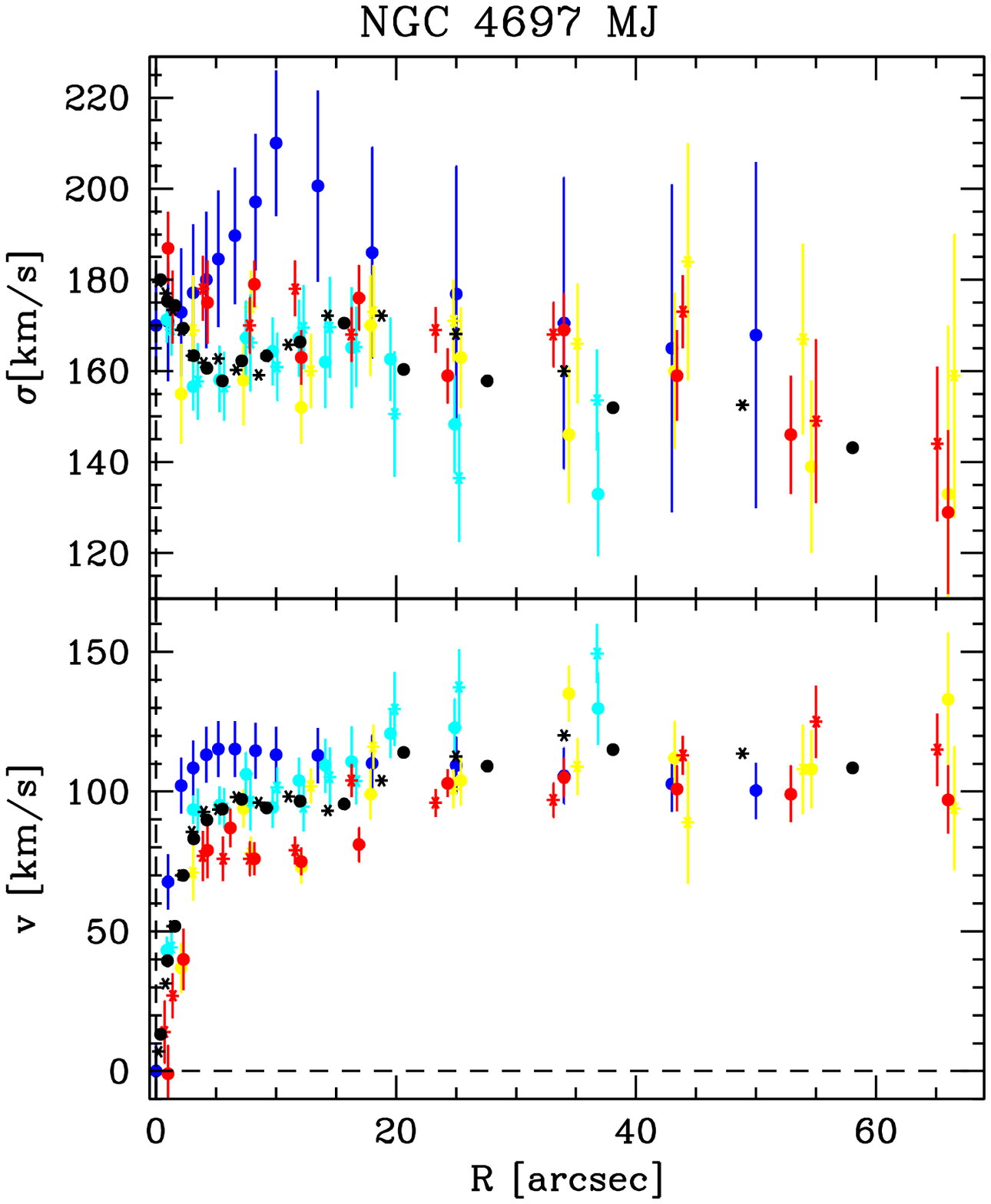}
\includegraphics[width=0.49\hsize,angle=0.0]{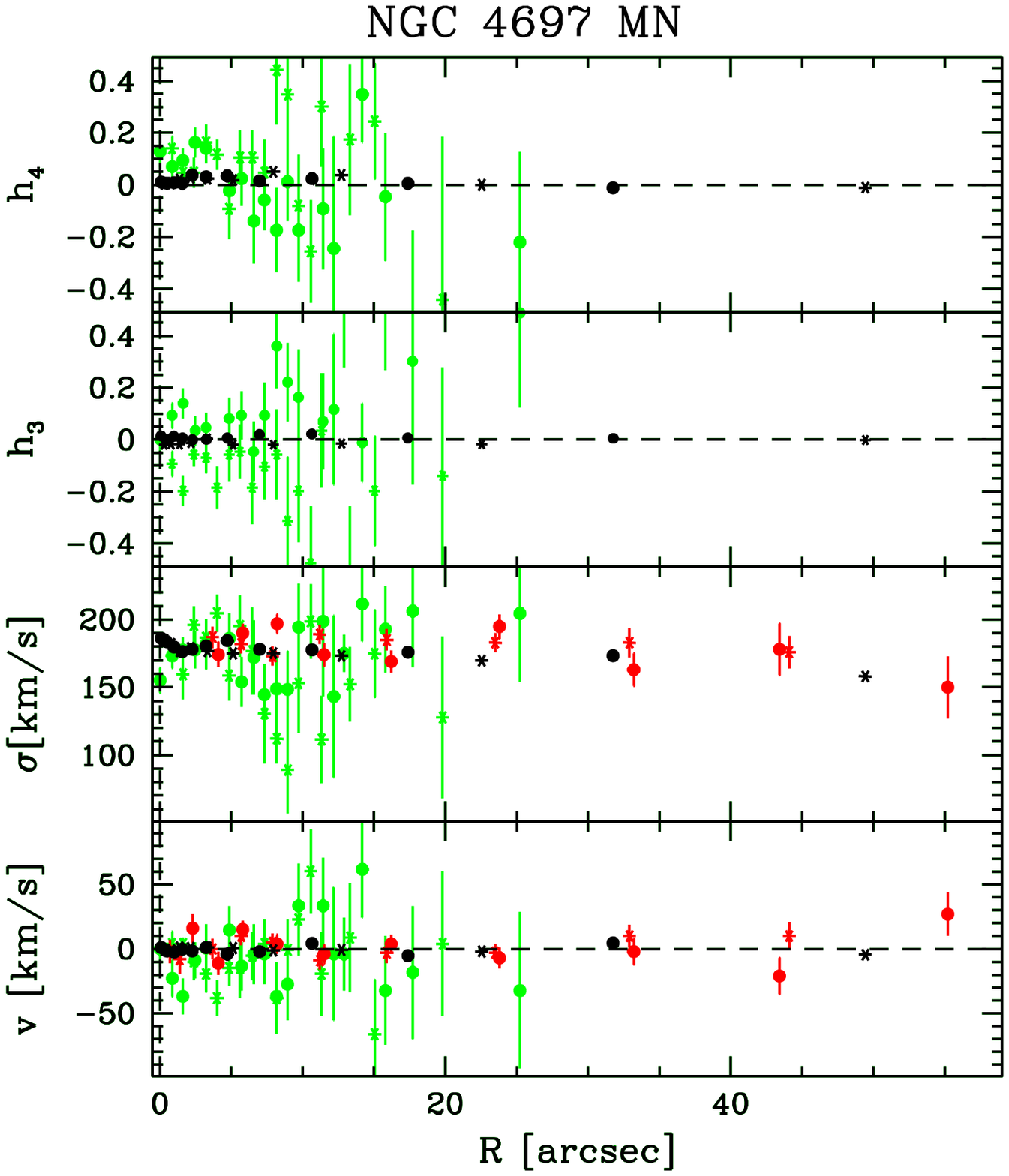}
\vskip0.5truecm
\caption[]{Comparison between the different absorption line kinematics
  along the major (left panel) and the minor (right panel) axis.
  Black: our data; red: BDI; blue: DDVZ; cyan: SB; green: KZ; yellow: FI.}
\label{fig:compall}
\end{figure*}

Along the major and minor axes we have derived additional
line-of-sight velocity distribution (LOSVD) kinematics from the high
S/N integrated absorption line spectra obtained by \citet{mendez+05}
with FORS2 at the VLT, a slit width of 1 arcsec and seeing
$1\arcsec-1.\arcsec5$. We refer to this paper for a description of the
data acquisition and reduction.  The LOSVDs were measured using the
Fourier Correlation Quotient (FCQ) method, as in \citet{BSG94} and
\citet{Mehlert+00}, and the K3III star HD132345 as a template.  From
these LOSVDs, profiles of $v$, $\sigma$, $h_3$ and $h_4$, the
Gauss-Hermite coefficients \citep{gerhard93,vdmarel_franx93}, were
obtained; these are shown in Figure~\ref{fig:kindata}. Tables
\ref{tabn4697mj} and \ref{tabn4697mn} give the data in tabular form.
The statistical errors derived from Monte Carlo simulations are minute
and much smaller than the rms scatter observed between the two sides
of the galaxy. These differences are particularly obvious along the
major axis in the radial range 10-20 arcsec. As noted in
\citet{mendez+05}, in this region we detect patchy [OIII] emission
that is affecting to some extent the kinematics.  Judging from the
asymmetries in the kinematics on both sides of the galaxy, we estimate
the residual systematic errors affecting the data, which amount to
$\approx 3$ km/s in $V$, $\approx 3.5$ km/s in $\sigma$, $\approx
0.02$ in $h_3$ and $h_4$.

In the following we discuss the comparison between the kinematic data
derived here and the kinematics published in the literature. Note that
part of the differences seen below arise from the different
observational setups (along the major axis different slit widths probe
different relative amounts of the central disk structure present in
the galaxy and related [OIII] emission regions) and methods used. In
particular, the FQ method fits Gaussian profiles to the LOSVDs,
ignoring the higher order Gauss-Hermite terms. In general the
systematic effect on the measured mean velocity and velocity
dispersion profiles is small \citep{vdm1994}. When applied to our
major axis data set, FQ gives systematically slightly smaller mean
velocities and larger velocity dispersions.

In the inner 10 arcsec along the major axis we confirm the clear
kinematic signature of the central disk discussed by
\citet[][SB]{scorza+95} and agree well with their mean velocities and
velocity dispersions derived also with the FCQ method and 1.8 arcsec
slit width (see Figure \ref{fig:compall}, left, cyan points). Along
the same axis we find good agreement with \citet{FI94} (FI, yellow,
derived using the FQ method and 1.1 arcsec slit width).  Overall, the
BDI data (Figure \ref{fig:compall}, red points, 2.6 arcsec slit
width), agree well with our data, although at $15\arcsec$ along the
major axis, the two data sets differ systematically.  The $\sigma$
profiles of DDVZ (blue points, 0.7 arcsec slit width) differ
significantly in the sense that at small semi-major axis distances the
DDVZ $\sigma$ is increasing with radius but our $\sigma$ is
decreasing.  Finally, the right panel of Figure~\ref{fig:compall}
compares our data along the minor axis with the data sets of BDI (red
points) and \citet{KZ00} (KZ, green points, 2 arcsec slit width), who
use the Fourier Fitting method of \citet{vdmarel_franx93}.  Both agree
within their respective (larger) errorbars.

Based on the radial extent and quality of the different datasets, and
taking into account the discussion above, we have decided to use in
the subsequent modeling only our data combined with BDI.  For our
kinematic data, the errors in $v$ and $\sigma$ are $\lta 0.5 \kms$,
which is small compared to the scatter in the data.  This suggests, as
already discussed above, that systematic errors dominate.  For the
modelling we have therefore replaced these errors with the smallest
errors in $v$ and $\sigma$ of the BDI data along the major and minor
axis, respectively ($5-7 \kms$). Similar arguments hold for $h_3$ and
$h_4$ and we set their errors to $0.01$. In addition, the $h_3$
coefficients along the minor axis scatter significantly around zero
while the minor axis velocities are consistent with zero; thus we
replace these $h_3$ values by $h_3=0.0$.

\begin{figure}
\centering 
\includegraphics[width=0.95\hsize,angle=-90.0]{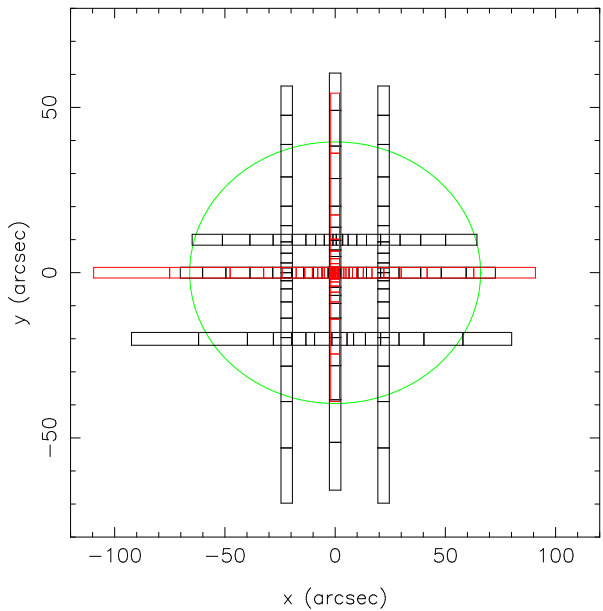}
\vskip0.5truecm
\caption[]{Schematic view of the slit setup used to construct the
  particle models. BDI slits are shown in black and our slits
  are shown in red.  The ellipse has a semi-major axis of length
  $R_{e}$ and axis ratio $q=0.6$.}
\label{fig:slits}
\end{figure}

Figure \ref{fig:slits} gives a schematic view of the arrangement of
the kinematic slits used in the modelling process.

\subsubsection{Planetary nebula velocities}
\label{sec:pnevs}
Planetary nebulae (PNe) are dying stars that emit most of their light
in a few narrow lines of which the $[\mathrm{OIII}]\lambda 5007$ is
the most prominent one. The PN population in elliptical galaxies is
expected to arise from the underlying galactic population of old stars
and hence the PNe can be used as kinematic tracers for the stellar
distribution. \citet{mendez_etal01} detected $535$ PNe in NGC 4697
and were able to measure radial velocities for $531$ of these
with a typical error of $40\;\kms$.  

\begin{figure}
\centering 
\includegraphics[width=0.7\hsize,angle=-90.0]{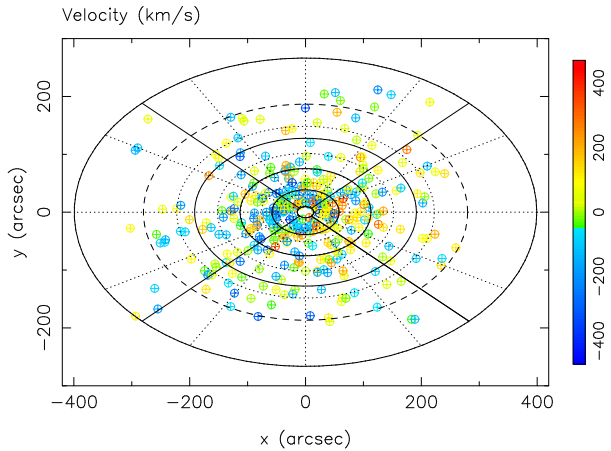}
\vskip0.5truecm
\caption[]{The positions and velocities of the cleaned PN sample of
  $381$ PNe. The lines indicate the different grids used for binning
  the PNe. Details are given in the text.}
\label{fig:pnedat}
\end{figure}

\citet{niri_etal06} analyzed the correlations between the magnitudes,
velocities and positions of these $531$ PNe and found kinematic
evidence for more than one PN sub-population in NGC 4697.  In addition
to the main PN population, they found evidence for a population of
preferably bright PNe which appeared to be not in dynamical
equilibrium in the galactic potential. To remove these possible
kinematic contaminants, and to also ensure completeness for $R > R_e$
\citep{mendez_etal01}, we discard all PNe with magnitudes outside the
range $26.2<m(5007)\le 27.6$. The positions and velocities of the
remaining $381$ PNe are shown in Figure \ref{fig:pnedat}.  In the
following, we use a doubled sample of $762$ PNe for our analysis,
obtained by applying a point-symmetry reflection.  Every PN with
observed position coordinates $(x,y)$ on the sky and line-of-sight
velocity $v_{PN}$ is reflected with respect to the center of the
galaxy to generate a new PN with coordinates $(-x,-y,-v_{PN})$.  Such
point-symmetric velocity fields are expected for axisymmetric and
non-rotating triaxial potentials.  Moreover, this reflection will help
to further reduce any PN sub-population biases which might still be
present.

We compute $v$ and $\sigma$ on two slightly different spatial grids,
subtracting $40\;\kms$ in quadrature from all PN velocity dispersions
to account for the measurement uncertainties \citep{mendez_etal01}.
We use the spatial bins defined by the solid lines displayed in
Figure~\ref{fig:pnedat} to obtain data set PND1, which is shown
together with the models in Section~\ref{sec:models}. The second data
set, PND2, is computed using the same grid but replacing the outermost
ellipse by the dashed ellipse with semi-major axis $a=280\arcsec$.
This second grid is used to make sure that the dynamical models we
generate are not affected by the way we define the outermost bins.

%
\section{NMAGIC models}
\label{sec:nmagic}
In this section we give a brief introduction to NMAGIC and present a
few extensions to the method described in \citet{delo+07}. \citet{ST96}
invented a particle-based method for constructing models of stellar
systems. This ``made-to-measure'' (M2M) method works by adjusting
individual particle weights as the model evolves, until the N-particle
system reproduces a set of target constraints. \citet{delo+07} improved
the algorithm to account for observational errors and to assess the
quality of a model for a set of target data directly, using the
standard $\chi^2$ statistics in the function to be maximized upon
convergence of the weights ($\chi^2$M2M). NMAGIC is a parallel
implementation of the improved $\chi^2$M2M algorithm.

\subsection{Luminous and dark matter distribution}

\subsubsection{Luminous mass}

We assume that the luminous mass distribution of NGC 4697 follows the
deprojected luminosity density. The mass density of the luminous matter is
then given by $\rho_\star=\Upsilon j$, with mass-to-light ratio
$\Upsilon$ and luminosity density $j$ represented by the discrete
ensemble of particles with positions ${\mathbf x}_i$ and luminosities
$L_i$.

\subsubsection{Dark halo potential}

The prevailing cosmological paradigm predicts that galaxies have
massive, extended dark matter halos.  Numerical cold dark matter (CDM)
simulations reveal universal halo density profiles with steep central
density cusps \citep[\eg][]{navarro+96,moore+99}. On the other hand,
observations of many dwarf and low-surface brightness galaxies find
shallower inner density cores \citep[\eg][]{deblok+03,mcgaugh+07}. Here
our aim is not to determine the detailed shape of the dark matter halo
in NGC 4697, but rather to first see whether the PN velocities allow
or require any dark matter at all in this galaxy.  To answer this
question we will investigate a one-dimensional sequence of potentials
whose circular velocity curves vary at large radii between the
near-Keplerian decline expected when the mass in stars dominates, and
the nearly flat shapes generated by massive dark halos. Thus for our
dynamical studies of NGC 4697, we represent the dark matter halo by
the logarithmic potential \citep{bin_tre87}
\begin{equation}
\phi_D(R',z') = \frac{v_0^2}{2}\ln(r_0^2+R'^2+\frac{z'^2}{q_\phi^2}),
\label{eqn:logpot}
\end{equation}
which is generated by the density distribution
\begin{equation}
\rho_D(R',z') = \frac{v_0^2}{4 \pi G q_\phi^2}
\frac{(2q_\phi^2+1)r_0^2+R'^2+2(1-\frac{1}{2}q_\phi^{-2})z'^2}
     {(r_0^2+R'^2+q_\phi^{-2}z'^2)^2},
\label{eqn:logdens}
\end{equation}
where $v_0$ and $r_0$ are constants, $q_\phi$ is the flattening of the
potential, and $R'$ and $z'$ are cylindrical coordinates with respect
to the halo's equatorial plane. When $q_\phi<1/\sqrt{2}$ the density
becomes negative along the $z'$ axis. The density given in equation
\ref{eqn:logdens} has a shallow inner density profile, but since we
are mainly interested in the circular velocity curve in the 
outer halo of NGC 4697, this is inconsequential: it is always possible
to reduce the stellar mass-to-light ratio in exchange for an additional
centrally concentrated dark matter cusp.
\subsubsection{The total gravitational potential}
\label{sec:globpot}

The total gravitational potential is generated by the combined
luminous mass and dark matter distributions and is given by
\begin{equation}
\phi = \phi_{\star}+\phi_D,
\label{eqn:totpot}
\end{equation}
where $\phi_{\star}$ is generated by the $N$-particle system assuming
a constant mass-to-light ratio for each stellar particle.  We estimate
$\phi_{\star}$ via a spherical harmonic decomposition
\citep{sellwood03,delo+07}.  The stellar potential is allowed to
change during a NMAGIC modelling run, but the dark matter potential is
constant in time and is given by equation~(\ref{eqn:logpot}).  The
particles are integrated in the global potential using a
drift-kick-drift form of the leapfrog scheme with a fixed time step.

\subsection{Model observables}
Typical model observables are surface or volume densities and
line-of-sight kinematics. An observable $y_j$ of a particle model is
computed via
\begin{equation}
y_j(t)=\sum_{i=1}^N w_i K_j \left[ {\mathbf z}_i(t) \right],
\label{eqn:obs}
\end{equation}
where $w_i$ are the particle weights, ${\mathbf z}_i$ are the
phase-space coordinates of the particles, $i=1,\cdots,N$, and
$K_j[{\mathbf z}_i(t)]$ is a kernel corresponding to $y_j$.  We use
units such that the luminosity $L_i$ of a stellar particle can be
written as $L_i = L w_i$, where $L$ is the total luminosity of the
model galaxy.  We use temporally smoothed observables to increase the
effective number of particles in the system, \cf \citet{ST96,delo+07}.

\subsubsection{Luminosity constraints}
\label{sec:alms}
For modeling the luminosity distribution of NGC 4697 one
can use as observables the surface density or space density on various
grids, or some functional representations of these densities. We have
chosen to model a spherical harmonics expansion of the deprojected
luminosity density.  We determine the expansion coefficients $A_{lm}$
for the target galaxy on a 1-D radial mesh of radii $r_k$. The
spherical harmonic coefficients for the particle model are computed
via
\begin{equation}
a_{lm,k} = L \sum_i \gamma_{ki}^{CIC} Y_l^m(\theta_i,\varphi_i) w_i,
\label{eqn:alm}
\end{equation}
where $L$ is the total luminosity of the model galaxy, $w_i$ the
particle weights, $Y_l^m$ the spherical harmonic functions and
$\gamma_{ki}^{CIC}$ is a selection function, which maps the particles
onto the radial mesh using a cloud-in-cell scheme
\citep[see][]{delo+07}.

\subsubsection{Kinematic constraints}
Since in the $\chi^2$M2M algorithm the kernel in
equation~(\ref{eqn:obs}) cannot depend on the particle weights
themselves, this puts some constraints on which observables can be
used. For kinematics, suitable observables are the luminosity-weighted
Gauss-Hermite coefficients or the luminosity-weighted velocity
moments. We implement them as follows.

\subparagraph*{Spectroscopic data}
The shape of the line-of-sight velocity distribution (LOSVD) can be
expressed as a truncated Gauss-Hermite series and is then
characterized by $V$, $\sigma$ and $h_n$ ($n>2$), where $V$ and
$\sigma$ are free parameters. If $V$ and $\sigma$ are equal to the
parameters of the best fitting Gaussian to the LOSVD, then $h_1=h_2=0$
\citep{vdmarel_franx93,rix_etal97}. The luminosity-weighted
Gauss-Hermite coefficients are computed as
\begin{equation}
b_{n,p} \equiv l_p\, h_{n,p} = 2\sqrt{\pi}L\sum_i \delta_{pi}
u_n(\nu_{pi}) w_i,
\label{eqn:lumghn}
\end{equation}
with
\begin{equation}
\nu_{pi}=\left(v_{z,i}-V_p\right)/\sigma_p.
\end{equation}
Here $v_{z,i}$ denotes the line-of-sight velocity of particle $i$,
$l_p$ is the luminosity in cell $\mathcal{C}_p$, $V_p$ and $\sigma_p$
are the best-fitting Gaussian parameters of the target LOSVD in cell
$\mathcal{C}_p$, and the dimensionless Gauss-Hermite functions are
\citep{gerhard93}
\begin{equation}
u_n(\nu) = \left(2^{n+1}\pi n!\right)^{-1/2}H_n(\nu)\exp \left(-\nu^2/2\right).
\end{equation} 
$H_n$ are the standard Hermite polynomials and $\delta_{pi}$ is a
selection function which is one if particle $i$ is in cell
$\mathcal{C}_p$ and zero otherwise.  The errors in $h_1$ and $h_2$ can
be computed from those of $V$ and $\sigma$ via
\begin{equation}
\Delta h_1 = -\frac{1}{\sqrt{2}} \frac{\Delta V}{\sigma}
\label{eqn:relh1}
\end{equation} 
and
\begin{equation}
\Delta h_2 = -\frac{1}{\sqrt{2}} \frac{\Delta \sigma}{\sigma},
\label{eqn:relh2}
\end{equation} 
valid to first order \citep{vdmarel_franx93,rix_etal97}.  Since we use
the observed $V_p$ and $\sigma_p$ from a Gauss-Hermite fit to the
LOSVD as expansion parameters for the model line profiles, the final
fitted $h_1$ and $h_2$ of a model will be small, and so we can also
use relations~(\ref{eqn:relh1}) and (\ref{eqn:relh2}) to compute the
model $V$ and $\sigma$ from $V_p$ and $\sigma_p$.
\subparagraph*{Spatially binned PNe data}
\label{sec:pne-binned-obs}
We have computed mean PN velocities and velocity dispersions for the
ellipse sector bins shown in Figure \ref{fig:pnedat}. The ellipticity
of the grid corresponds to the mean ellipticity of the photometry.  As
suitable observables we take the {\it luminosity-weighted} velocity
moments in these bins, which are computed as
\begin{equation}
v^n_p  = L\sum_i \delta_{pi} v^n_{z,i} w_i, 
\end{equation}
where $v_{z,i}$ is the velocity along the line-of-sight of particle
$i$ and $\delta_{pi}$ is a selection function, which is equal to one
if particle $i$ belongs to the bin segment under consideration and
zero otherwise. In the following, we use only the moments $v_p^1$
and $v_p^2$.

\subsection{Seeing effects}
\label{sec:seeing}
To account for seeing effects we apply a Monte Carlo approach
\citep[\eg][]{cappellari+06} instead of convolving the observables
with the PSF. As long as the particles move along their orbits no PSF
effects need to be taken into account, only when the observables of
the system are computed, the effects of seeing may matter.

When computing an observable including PSF effects, we replace the
``original'' particle at position $(x_i,y_i)$ on the sky plane
temporarily by a cloud of $N_{pp}$ pseudo particles. The position of a
pseudo particle is obtained by randomly perturbing $(x_i,y_i)$ with
probability given by the PSF. Note that neither extra storage is
needed nor additional time to integrate the particles along their
orbits. Usually, only a small number of pseudo particles are needed to
model PSF effects, even one is often sufficient.  This procedure is
implemented in the kernel $K_{ij}$ as defined in
equation~(\ref{eqn:obs}). The same kernel then enters the
force-of-change equation, \cf \cite{delo+07}.

\begin{figure}
\centering 
\includegraphics[angle=-90.0,width=0.95\hsize]{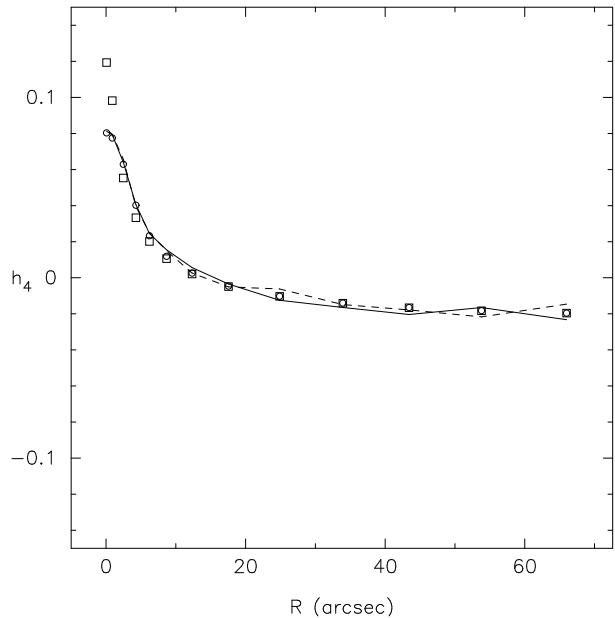}
\vskip0.2truecm
\caption[]{Seeing convolution test, comparing the radial run of $h_4$
  along a $2\arcsec$ slit for a spherical target model and its
  particle representation. The squares (circles) were computed for the
  target from higher order Jeans equations without (with) seeing. The
  lines correspond to the particle model including seeing, for which
  the PSF was represented using $N_{pp}=5$. The full and dashed lines
  refer to the major axis slit data at positive resp.\ negative radii
  with respect to the origin. The heavily seeing-affected central
  profile is well recovered by the model.}
\label{fig:seeintest}
\end{figure}

To test how well PSF effects are modeled using only a few pseudo
particles we computed mock observations for a spherical isotropic
galaxy of mass $M=10^{10} M_\odot$, located at a distance
$10\;\mathrm{Mpc}$. The intrinsic density of the galaxy is given by a
\citet{hernquist90} profile with scale length $a=55.1\arcsec$. We
assumed a major axis slit of width $2\arcsec$ and a Gaussian PSF with
$\mathrm{FWHM} = 4 \arcsec$.  We computed the LOSVD of the target
galaxy along the major axis using higher order Jeans moments
\citep{magorrian_bin94}, and compared it with the LOSVD of a particle
realization of the mock galaxy, applying the above procedure. The
particle realization of the Hernquist model was generated from an
isotropic distribution function \citep[\eg][]{victor_sell00}. As an
example, Figure~\ref{fig:seeintest} shows the $h_4$ profile along the
major axis slit. The square symbols are the target data computed using
higher order Jeans equations without seeing, the circles are computed
the same way but including PSF effects. The lines are the temporally
smoothed $h_4$ profiles of the particle model using $N_{pp}=5$ to
represent the PSF. One sees that the heavily seeing-affected central
profile is well recovered by the model.

In the dynamical modeling of NGC 4697 we include seeing only for our
new kinematic data. We represent the PSF by a single Gaussian with $FWHM
= 1.25\arcsec$. For the BDI data we do not know the PSF but since the
slit cells are relatively large, seeing is likely to be negligible.

\subsection{The merit function}
\label{sec:entropy}
By fitting the particle model to the observables, the weights $w_i$
are gradually changed such that the merit function
\begin{equation}
F = \mu S-\frac{1}{2}\chi^2
\label{eqn:F}
\end{equation}
is maximized, where $S$ is a profit function and $\chi^2$ measures the
quality of the fit. The parameter $\mu$ controls the relative
contribution of the profit function to $F$; incrementing $\mu$
increases the influence of S in equation (\ref{eqn:F}).  The $\chi^2$
statistics is computed as usual
\begin{equation}
\chi^2  = \sum_j \Delta_j^2,
\label{eqn:chi2}
\end{equation}
where $\Delta_j=(y_j-Y_j)/\sigma(Y_j)$. $y_j$ is a model observable
(\eg $a_{lm,k}$ with ${\mathbf j} =\{lm,k\}$), $Y_j$ is the
corresponding target and $\sigma(Y_j)$ its error.

For the profit function $S$, we use the entropy
\begin{equation}
S = -\sum_i w_i \ln (w_i/\hat{w}_i)
\label{eqn:entropy}
\end{equation}
where $\{\hat{w}_i\}$ are a predetermined set of weights, the
so-called priors (here equal for all particles). The entropy term
pushes the particle weights to remain close to their priors (more
specifically, close to $\hat{w}_i/e$). This implies that models with
large $\mu$ will have smoother distribution functions than those with
small $\mu$. The best choice for $\mu$ depends on the observational
data to be modeled, \eg their spatial coverage, on the phase-space
structure of the galaxy under consideration, but also on the initial
conditions. For the dataset at hand, the best value of $\mu$ will
be determined in Section \ref{sec:mu}
\subsection{Discrete PNe velocities}
\label{sec:pnedisc}
The likelihood of a model fit to photometric as well absorption line
kinematic data is measured by the standard $\chi^2$ statistics given
in equation~(\ref{eqn:chi2}).  To treat discrete PN velocity
measurements the same way, we must bin them to estimate the underlying
mean $v$ and $\sigma$ fields. This gives the corresponding model
observables as discussed in section \ref{sec:pne-binned-obs}.

As an alternative, one can measure the likelihood of a sample of
discrete velocities $v_j$ and positions $\mathbf{R}_j=(x_j,y_j)$ 
on the sky via
\begin{equation}
  \label{eqn:LHtot}
  {\mathcal L} = \sum_j \ln {\mathcal L}_j
\end{equation}
using the likelihood function for a single PN \citep{romano_koch01}
\begin{equation}
  \label{eqn:LH}
  {\mathcal L}_j(v_j,\mathbf{R}_j)=\frac{1}{\sqrt{2\pi}}\int 
  \frac{\ud L}{\ud v_z}(v_z,\mathbf{R}_j) e^{-(v_j-v_z)^2/2\sigma_j^2}
      \ud v_z,
\end{equation}
where $\sigma_j$ is the error in velocity and $\ud L/\ud v_z$ is the
LOSVD assuming as before that the line-of-sight is along the z-axis.

We can then add equation (\ref{eqn:LHtot}) to the function F given in
equation (\ref{eqn:F}) and maximize
\begin{equation}
F^+ = F+{\mathcal L}
\label{eqn:FLH}
\end{equation}
with respect to the particle weights $w_i$. Hence, we obtain an
additional contribution to the force-of-change as given in
\citet{delo+07}. We will now derive this extra term.  Let us consider
the selection function
\[ \delta_{ji} = \left \{ \begin{array}{ll}
1 & \mbox{if $(x_i,y_i) \in \mathcal{C}_j$} \\
0 & \mbox{otherwise.}
	\end{array}
\right. \]
which assigns particle weights to a spatial cell
$\mathcal{C}_j$, which contains the $j$-th PNe.
We can then write $\ud L/\ud v_z$ at position $j$ as
\begin{equation}
\left(\frac{\ud L}{\ud v_z}\right)_j = \frac{1}{l_j}\sum_i \delta_{ji} w_i \delta(v_z-v_{z,i})
\label{eqn:LOSVD}
\end{equation}
with
\begin{equation}
l_j = \sum_i \delta_{ji} w_i,
\label{eqn:LOSVDnorm}
\end{equation}
and $\delta(x)$ being the standard delta function. Hence, equation (\ref{eqn:LH})
can be expressed in terms of the particles via
\begin{equation}
{\mathcal L}_j = \frac{\widehat{{\mathcal L}}_j}{l_j}
\end{equation}
with
\begin{equation}
\widehat{{\mathcal L}}_j = \frac{1}{\sqrt{2\pi}} \sum_i \delta_{ji} w_i e^{-(v_j-v_{z,i})^2/2\sigma_j^2}.
\end{equation}
Finally, we find for the additional term in the FOC
\begin{equation}
\frac{\ud w_i}{\ud t} = \varepsilon w_i \sum_j \delta_{ji}
\left(\frac{1}{\sqrt{2\pi}}\frac{e^{-(v_j-v_{z,i})^2/2\sigma_j^2}}{\widehat{{\mathcal
L}}_j} - \frac{1}{l_j} \right),
\label{eqn:FOC+}
\end{equation}
where the sum runs over all individual PNe. For small errors, the $\ud
w_i/\ud t$ from the likelihood term is positive for particles with
$v_j=v_{z,i}$, but reduces the weights of the other particles and
hence drives the LOSVD to peak at $v_j$.  In the implementation, we
replace $l_j$ and $\widehat{{\mathcal L}}_j$ with the corresponding
temporally smoothed quantities.

When we use this method to account for the PN velocities in NGC 4697,
we adopt the grid defined in Figure \ref{fig:pnedat} by the dotted
lines, including the innermost and outermost full ellipses. In this
way, we assign each of the $762$ PNe to a cell $\mathcal{C}_j$.  It
follows, that more than one PNe share the same spatial bin, but this
is not a problem.

\subsection{Efficient mass-to-light estimate}
\label{sec:FOCM2L}
It is common practice to evolve N-particle systems in internal units
(IU), in which the gravitational constant and the units of length
and mass are set to unity, and to scale the system to physical
units (PU) a posteriori to compare with galaxy observations.
Similarly, the velocities of a system with mass-to-light ratio
$\Upsilon$ of unity may be scaled to any $\Upsilon$ via $v_{\rm PU} =
\gamma \; v_{\rm IU}$ where $\gamma \propto \sqrt{\Upsilon}$ and $v_{\rm
  PU}$ and $v_{\rm IU}$ are the velocities in physical and internal
units, respectively. It follows that the kinematic observables of the
model and hence also $\chi^2$ can be regarded as functions of
$\Upsilon$. Equation (\ref{eqn:chi2}) then reads
\begin{equation}
  \label{eqn:chi2ups}
  \chi^2 = \sum_j \Delta_j(\Upsilon)^2.
\end{equation}
In the following we will only consider the luminosity-weighted
Gauss-Hermite moments as given in equation (\ref{eqn:lumghn}) and
neglect the PNe kinematic constraints.  Taking the partial derivative
with respect to $\Upsilon$ of equation (\ref{eqn:chi2ups}) leads to
\begin{equation}
  \label{eqn:dchi2ups}
 \frac{1}{2}\frac{ \partial \chi^2 }{\partial \Upsilon} = 
 \sum_{\bf j} \frac{\Delta_j(\Upsilon)}{\sigma(B_{n,p})}
 \frac{ \partial b_{n,p} }{\partial \Upsilon},
 \,\,\, {\mathbf j}=\{n,p\}
\end{equation}
where $B_{n,p}$ is the target observable and $\sigma(B_{n,p})$ its error. 
We define a force-of-change (FOC) for the mass-to-light ratio
$\Upsilon$ 
\begin{equation}
 \frac{ \ud \Upsilon }{\ud t} = -\eta \Upsilon 
 \frac{ \partial \chi^2 }{\partial \Upsilon}
\end{equation}
which equals
\begin{equation}
\label{eqn:FOCM2L}
 \frac{ \ud \Upsilon }{\ud t} = -\eta \Upsilon
 \sum_{\bf j} 2 \Delta_j(\Upsilon) 
          \frac{\partial \Delta_j(\Upsilon)}{\partial \Upsilon}
\end{equation}
with
\begin{equation}
 \frac{\partial \Delta_j(\Upsilon)}{\partial \Upsilon}
 = \frac{\sqrt{\pi} L}{\Upsilon \sigma_p\sigma(B_{n,p})}
 \sum_{i} \delta_{p i} w_i \frac{\partial u_n(x)}{\partial x} \Big
 |_{x=\nu_{pi}} v_{z,i},
\end{equation}
where we used $\partial v_{z,i}/\partial \Upsilon = v_{z,i}/2\Upsilon$
for $v_{z,i}$ given in physical units, and ${\mathbf j}=\{n,p\}$. The
line-of-sight is along the z axis. In practice, we use the temporally
smoothed quantities to compute the FOC for  the mass-to-light ratio.

In principle, the proposed scheme can be understood as a gradient
search along the $\chi^2(\Upsilon)$ curve when simultaneously the
particle model is fitted to the observational constraints. Hence the
same NMAGIC run allows us to estimate $\Upsilon$ as well. We test the
scheme and illustrate its accuracy in Section \ref{sec:isorot}.
\subsection{Initial conditions}
\label{sec:ics}
As initial conditions for NMAGIC, we generate a particle realization
of a spherical $\gamma$-model \citep{dehnen93,carollo_etal95} made
from a distribution function (DF) using the method of
\citet{victor_sell00}.  The model consists of $N=5\times 10^5$
particles and has $\gamma=1.5$, scale length $a=1$ and $r_{\rm
  max}=40$. When scaled to NGC 4697 one unit of length corresponds to
$2.3810\;\mathrm{kpc}$, i.e., this model has $R_e=3.8\;\mathrm{kpc}$.

In some cases, we have found it useful to give the initial particle
system some angular momentum about an axis of symmetry. For
axisymmetric stellar systems, the density is determined
through the even part in $L_z$ of the DF \citep{lyndenbell62}. Thus
the component of the angular momentum of a particle along the symmetry
axis may be reversed without affecting the equilibrium of the system.
\citet{kalnajs77} showed, however, that a discontinuity at $L_z=0$ can
affect the stability of the particle model. Therefore, if desired, we
switch retrograde particles with a probability
\begin{equation}
p(L_z) = p_0 \frac{L_z^2}{L_z^2+L_0^2},
\label{eqn:prob}
\end{equation}
which ensures a smooth DF. 

\section{Testing the modelling with isotropic rotator targets}
\label{sec:isorot}
In this section, we use axisymmetric, isotropic rotator models with
known intrinsic properties to determine the optimal value of the
entropy ``smoothing'' parameter $\mu$ in equation (\ref{eqn:F}), and
to test our procedure for determining the optimal mass-to-light ratio
simultaneously with modelling the data.

\subsection{Entropy parameter $\mu$}
\label{sec:mu}

\begin{figure}
\centering 
\includegraphics[angle=-90.0,width=0.95\hsize]{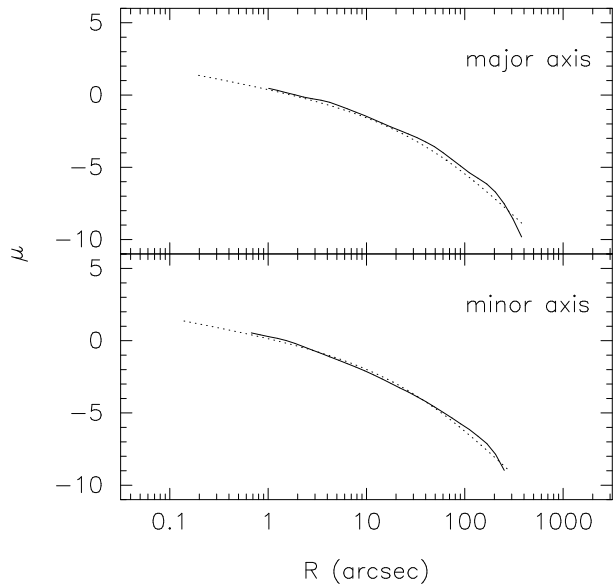}
\vskip0.2truecm
\caption[]{Comparison of the surface brightness of NGC 4697 (solid
  lines) with the $\gamma$-model described in the text and seen under
  $i=80^\circ$ inclination (dotted lines). Top: Surface brightness
  profile along the major axis.  (b) Along the minor axis.}
\label{fig:gamma_vs_n4697}
\end{figure}

Our approach to determine suitable values for $\mu$ is similar as in
\citet{gerhard_etal98,thomas_etal05}. We first generate a ``mock''
kinematic data set from an isotropic rotator model whose information
content (number and density of points, errors) is similar as for the
real data set to be modelled. To this data set we perform a sequence
of particle model fits for various $\mu$, and determine the values of
$\mu$ for which (i) a good fit is obtained, and (ii) the known
intrinsic velocity moments of the input ``mock'' system are well
reproduced by the corresponding moments of the final particle model.
Using an isotropic rotator model for this purpose here makes sense,
because such a model is a fair representation of NGC 4697
\citep{binney_etal90}.  We have chosen to describe the luminosity
density of the mock galaxy by one of the flattened $\gamma$-models
of \citet{dehnen_ger94},
\begin{equation}
j(m) = \frac{(3-\gamma)L}{4\pi q}\frac{a}{m^\gamma(m+a)^{4-\gamma}}.
\label{eqn:lumgam}
\end{equation}
Here $L$ and $a$ are the total luminosity and scale radius, $m^2 =
R'^2+(z'/q)^2$, and $q$ is the flattening. The parameters are chosen
such that the surface brightness closely resembles that of NGC 4697,
i.e., $q=0.7$, $\gamma=1.5$, $L=2 \times 10^{10}\;L_{\odot,R}$ and
$a=2.5 \;\mathrm{kpc}$, which corresponds to $a\approx 49 ''$ at a
distance of $10.5\;\mathrm{Mpc}$. Figure \ref{fig:gamma_vs_n4697}
shows a comparison of the surface brightness of NGC 4697 with the mock
galaxy projected under $i=80^\circ$. The major and minor axis surface
brightness profiles are well approximated by the $\gamma$-model,
except for some differences at larger radii, so we will use this model
for the calibration of $\mu$.

\begin{figure}
\centering 
\includegraphics[angle=-90.0,width=0.95\hsize]{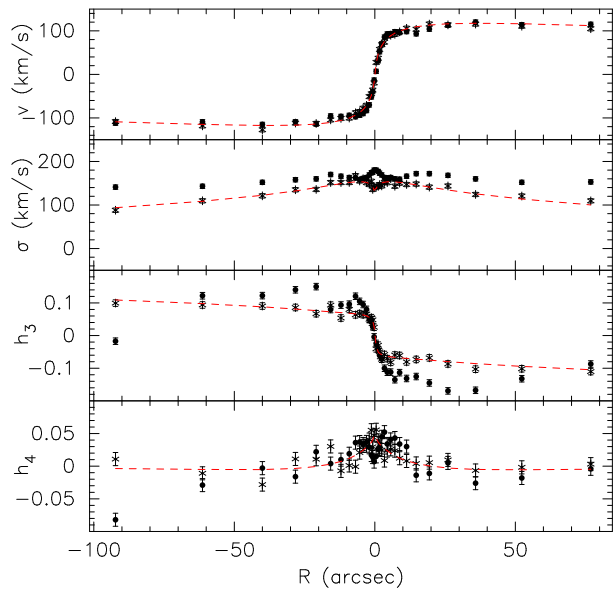}
\vskip0.2truecm
\caption[]{Comparison of $v$, $\sigma$, $h_3$ and $h_4$ of NGC 4697
  and an isotropic rotator model with approximately the same projected
  surface brightness as the galaxy. The filled circles show our new
  kinematic data for NGC 4697 from Section \ref{sec:kindata}, the star
  symbols show the isotropic rotator mock data, and the dashed red lines
  show the underlying smooth model kinematics, all along the major
  axis.}
\label{fig:S_mj_iso_vsn4697}
\end{figure}

We determine mock kinematic profiles from internal velocity moments,
obtained from higher-order Jeans equations \citep{magorrian_bin94} in
the self-consistent potential generated by the density of equation
(\ref{eqn:lumgam}) for a mass-to-light ratio $\Upsilon=5$.  Before
calculating the line-of-sight velocity profiles, the velocity moments
are slit-averaged to account for the observational setup of the
kinematic slits given in Section \ref{sec:kindata}. We add Gaussian
random variates to the isotropic rotator kinematics with $1\sigma$
dispersion corresponding to the respective measurement error in NGC
4697 at that position.  Figure \ref{fig:S_mj_iso_vsn4697} shows a
comparison of our new kinematic data for NGC 4697 with the isotropic
rotator mock data, along the galaxy's major axis.

We do not construct mock PNe data for inclusion in the entropy tests,
but we need to compute the photometric observables to construct a
complete observational data set. We expand the luminosity distribution
of equation (\ref{eqn:lumgam}) in a spherical harmonics series (\cf
Section \ref{sec:alms}) on a radial grid with 40 shells at radii
$r_k$. The radii are quasi-logarithmically spaced with $r_{\rm
  min}=1.0\arcsec$ and $r_{\rm max}=700\arcsec$. We use the luminosity
on radial shells $L_k=\sqrt{4\pi} A_{00,k}$ and the higher order
coefficients $A_{20,k}$, $A_{22,k}$, $\cdots$, $A_{66,k}$ and
$A_{80,k}$ to constrain the luminosity distribution of the particle
model. The $m\ne0$ terms are set to zero to force the models to remain
nearly axisymmetric, i.e., within the limits set by the $A_{lm}$
errors.  We assume Poisson errors for the $L_k$:
$\sigma(L_k)=\sqrt{L_k L/N}$ where N is the total number of particles
used in the particle model and $L$ is the total luminosity of the
system. To estimate the errors in the higher order luminosity moments,
we use Monte-Carlo experiments in which we generate particle
realizations of a spherical approximation of the density field of the
target system with $5 \times 10^5$ particles, which is the same number
as in the $\chi^2$M2M models for NGC 4697.

We then construct self-consistent particle models for the isotropic
rotator target in a three step process, using the mock data as
constraints. (i) Density fit: we start with the spherical initial
conditions described in Section \ref{sec:ics} and evolve them using
NMAGIC to generate a self-consistent particle realization with the
desired luminosity distribution ($\gamma$-particle model), fitting
only the luminosity constraints. (ii) Kinematic fit: because the
target galaxy has a fair amount of rotation, it is worth starting the
kinematic fit from a rotating model. Hence, following Section
\ref{sec:ics}, we switch a fraction of retrograde particles in the
$\gamma$-particle model to prograde orbits, using $p_0=0.3$ and
$L_0\simeq L_{\rm circ}(0.03 R_e)$. This rotating system we then use
as a starting point to construct a series of self-consistent dynamical
isotropic rotator models, by fitting the target photometry {\sl and}
kinematics for different values of $\mu$.  For each model, we evolve
the particle system for $\sim 10^5$ NMAGIC correction steps while
fitting the complete set of constraints. During this correction phase,
the potential generated by the particles is updated after each
correction step. (iii) Free evolution: to ensure that any correlations
which might have been generated during the correction phase are
phase-mixed away, we now keep the potential constant and evolve the
system freely for another $5000$ steps, without further correction
steps. For reference, $5000$ of these steps correspond to $\approx 20$
circular rotation periods (``dynamical times'')at $R_{e}$ in spherical
approximation.
%
\begin{figure}
\centering 
\includegraphics[angle=-90.0,width=0.95\hsize]{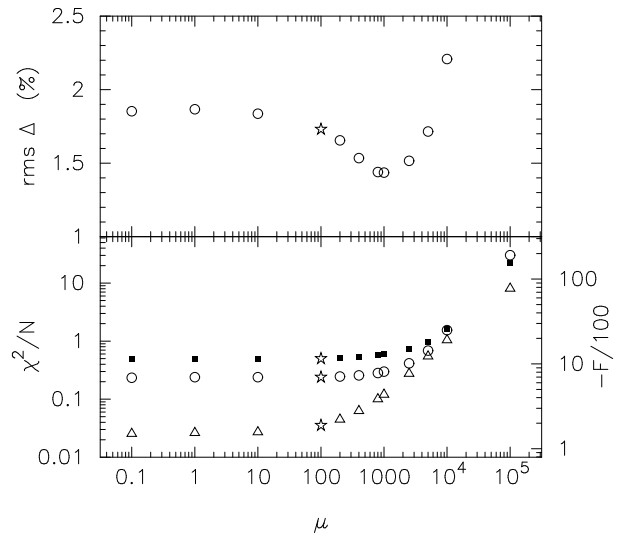}
\vskip0.2truecm
\caption[]{Entropy tests. Top: Deviation ${\rm rms}\;\Delta(\mu)$ of
  the particle models from the isotropic rotator internal velocity
  moments. The point for the rightmost value of $\mu$ is at a large
  value of $\Delta$ outside the diagram.  Bottom: $\chi^2$ deviation
  per data point of the particle model fit to the photometric and
  kinematic target observables (open circles) and to the kinematic
  observables alone (filled squares), as a function of entropy
  parameter $\mu$. The triangles show the same dependence for the
  merit function ($-F$), \cf equation (\ref{eqn:F}). The starred
  symbol indicates the value of $\mu$ chosen for the subsequent
  modelling.}
\label{fig:chi2mu}
\end{figure}

The results are presented in Figure \ref{fig:chi2mu}.  The lower panel
shows the quality of the fit as a function of $\mu$, both in terms of
normalized $\chi^2$ values and in terms of the merit function $F$ from
equation (\ref{eqn:F}). The upper panel shows the ${\rm rms}\; \Delta$
relative difference between the internal velocity moments of the
isotropic rotator input model and those of the particle models
reconstructed from the mock kinematics.  For the particle models
intrinsic velocity moments are computed by binning the particles in
spherical polar coordinates, using a quasi-logarithmic grid with 20
radial shells bounded by $r_{\rm min} = 0.01 \arcsec$ and $r_{\rm max}
= 200 \arcsec$, 12 bins in azimuthal angle $\phi$, and 21 bins equally
spaced in $\sin \theta$.  The ${\rm rms}\; \Delta$ shown in Figure
\ref{fig:chi2mu} is obtained by averaging over all grid points in the
radial region constrained by the data ($R\le 1.5 R_e$).  The minimum
in ${\rm rms}\; \Delta$ determines the value of $\mu$ for which the
model best recovers the internal moments of the input model. This
occurs at $\mu\simeq 10^{3}$, and the value of ${\rm rms}\; \Delta$ at
the minimum is $\simeq 1.4\%$ . For larger (smaller) $\mu$, the ${\rm
  rms}\; \Delta$ is larger because of oversmoothing (excess
fluctuations) in the model.

$\chi^2/N$ values are given in the lower panel of
Fig.~\ref{fig:chi2mu} for all (photometric and kinematic) data points,
and for the kinematic data points alone. Generally the $\chi^2/N$ for
the photometric points is significantly better than for the kinematic
points, because (i) the $A_{lm}$ come from averages over many
particles, thus have little noise, and we have not added Gaussian
variances, and (ii) all particles contribute to the $A_{lm}$
force-of-change at all timesteps, so the $A_{lm}$ are weighted
strongly during the evolution.  The kinematic $\chi^2$ per data point
in the lower panel is of order $0.5$ for a large range of $\mu$ and
then increases starting from $\mu \gta 300$ to $1$ at
$\mu\simeq5\times10^3$, whereas $-F$ already increases around $\mu
\gta 100$.

\begin{figure}
\centering
\includegraphics[angle=-90.0,width=0.95\hsize]{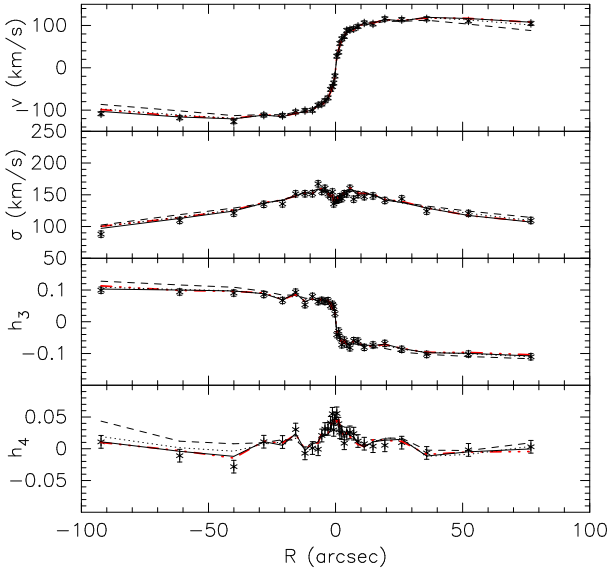}
\vskip0.2truecm
\caption[]{Particle model fits to the isotropic rotator mock kinematic
  data along the model's major axis. The points with error bars show
  the target data and the lines represent the model kinematics. The
  model data points are averages over the same slit cells as the
  target data (see Fig.~\ref{fig:slits}), and are connected by
  straight line segments.  The model $v$, $\sigma$ are determined via
  eqs.~(\ref{eqn:relh1}) and (\ref{eqn:relh2}) and $h_3$, $h_4$ are
  the fitted values based on the observed scale parameters $V_p$ and
  $\sigma_p$. The full, dotted and dashed lines correspond to the
  models obtained with $\mu=10^2$, $\mu=10^3$, and $\mu=5\times10^3$,
  respectively. The red dashed line shows the $\mu=100$ model
  kinematics $20$ dynamical times later, reflected with respect to the
  origin, and obtained from direct fitting of the model line profiles.
  This proves that this model is accurately axisymmetric and
  stationary; see text for details.}
\label{fig:isofit_mj}
\end{figure}

\begin{figure}
\centering 
\includegraphics[angle=-90.0,width=0.95\hsize]{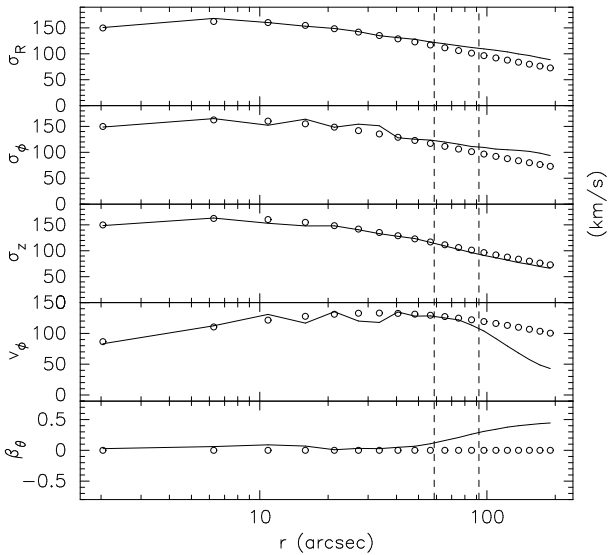}
\vskip0.2truecm
\caption[]{Comparison of the intrinsic velocity moments in the
  equatorial plane of the axisymmetric isotropic rotator particle
  model and target model. The points represent the target system and
  the lines correspond to the final particle model for $\mu=100$,
  averaged over azimuth. The dashed vertical lines show the maximum
  radial extent of the minor axis (left line) and major axis target
  kinematic data (right line).  At larger radii the particle model is
  poorly constrained by the input target data.}
\label{fig:intkin_maj}
\end{figure}

Some results for the isotropic rotator dynamical models obtained with
$\mu=10^2$, $\mu=10^3$, $\mu=5\times10^3$ are presented in Figures
\ref{fig:isofit_mj} and \ref{fig:intkin_maj}.
%
Figure \ref{fig:isofit_mj} shows a comparison of the target kinematics
with the kinematics of the self-consistent particle models along the
major axis slit.  Note the excellent fit of the central velocity
gradient and velocity dispersion dip, for all $\mu$ values.  However,
the models with higher $\mu$ begin to fail matching the target data at
the largest radii. This is because the number of data points decreases
with radius, whereas the number of particles and hence entropy
constraints is roughly proportional to luminosity $L_k$, i.e.\ changes
much more slowly with radius. The result is that the constraints from
the data become relatively weaker at larger radii. The entropy term
tries to enforce a dynamical structure related to the initial particle
model, in which all particles have equal weights. In the present case
this works to cause a bias against both fast rotation and anisotropy.
This first becomes apparent where the relative statistical power of
the data is weakest, i.e., at large radii.

Because our goal in this paper is to determine the range of potentials
in which we can find valid dynamical models for NGC 4697, we need to
ensure that the answer to this question is not biased by overly strong
entropy smoothing in the galaxy's outer regions. Thus in the modelling
in Section~\ref{sec:models} we will conservatively choose $\mu=100$
for the smoothing parameter (indicated by the starred symbol in
Fig.~\ref{fig:chi2mu}). Similar caution is common practice in
determining black hole masses in galaxies
\citep[e.g.,][]{gebhardt_etal03}. The resulting dynamical models will
then be somewhat less smooth than could be achieved, but this price is
rather modest; between $\mu=10^2$ and its minimum value at $\mu=10^3$,
the rms $\Delta$ in Fig.~\ref{fig:chi2mu} decreases from $\simeq
1.7\%$ to $\simeq 1.4\%$, i.e., by $\simeq 15\%$. Certainly it
would not be appropriate to rule out potentials in which the solutions
differ by this degree in smoothness.

Using $\mu=100$ in the modelling leads to a slight overfitting of the
slit kinematic data, especially for the higher order kinematic moments
which themselves take only values of order percents.  It is worth
pointing out that, contrary to first appearances from
Fig.~\ref{fig:isofit_mj}, this implies neither that these models are
not axisymmetric, not that they are out of equilibrium.  The model
kinematics shown in Fig.~\ref{fig:isofit_mj} are obtained after $20$
dynamical times of free evolution in the axisymmetric potential, so
are thoroughly phase-mixed.  The model data points shown are averages
over the same slit cells as the target data (see
Fig.~\ref{fig:slits}), and are connected by straight line segments.
The plotted $v$, $\sigma$ are determined via eqs.~(\ref{eqn:relh1})
and (\ref{eqn:relh2}) and $h_3$, $h_4$ are the fitted values based on
the observed scale parameters $V_p$ and $\sigma_p$. The red dashed
line in Fig.~\ref{fig:isofit_mj} also corresponds to the $\mu=100$
model, but has been determined as follows: (i) from the particle
distribution after a total of $40$ dynamical times of free evolution;
(ii) from a mirror-symmetric set of slit cells, with respect to the
major-axis slit shown in Fig.~\ref{fig:slits}, (iii) using the
$(v,\sigma,h_3,h_4)$ parameters obtained by direct fits to the model
line profiles, and (iv) finally reflecting the kinematics so obtained
anti-symmetrically with respect to the origin. The excellent agreement
between this curve and the original major axis kinematics of this
model in Fig.~\ref{fig:isofit_mj} shows that (i) the $\mu=100$ model
is a true equilibrium, (ii) it is accurately axisymmetric, and (iii)
the left-right differences in the kinematics in
Fig.~\ref{fig:isofit_mj} are due to slightly different slit cell
averages over the model on both sides. That these averages can be
slightly different is made possible by low-level (axi-symmetric)
structure in the model consistent with the slight under-smoothing for
this value of $\mu$.  What happens is that the algorithm adds a few
near-circular orbits in the relevant radial ranges. When added to the
corresponding model LOSVDs and averaged over the asymmmetric slit
cells these orbits change the kinematic moments $h_n$, $n\ge3$ at the
$\simeq0.01$ level so as to improve the agreement with the observed
major axis $h_3$, $h_4$.  In the other slits the models interpolate
more smoothly between points when needed because fluctuations in the
particle distribution to follow local kinematic features are less
easily arranged; see the corresponding figure for NGC 4697 in Section
\ref{sec:models}.

A comparison of the internal velocity moments of the input model and
the particle model in the equatorial plane is presented in
Figure~\ref{fig:intkin_maj}. The figure shows $\sigma_R$,
$\sigma_\phi$ and $\sigma_z$, followed by $v_\phi$. The last panel
displays the anisotropy parameter
$\beta_\theta=1-\sigma^2_\theta/\sigma^2_r$, which is zero for the
input isotropic rotator model.  Within the radial extent of the target
data, the internal moments of the input model are well reproduced;
outside this region, where the model is poorly constrained by the
input data, the particle model increasingly deviates from the target.
Indeed, if we add PN velocity data in this test, the corresponding
particle model agrees with the internal moments of the input model out
to larger radii.
%

\subsection{Mass-to-light ratio}
We will now use such isotropic rotator models to explore how
accurately we are able to recover the input mass-to-light ratio, given
the spatial coverage of the data. Further, we will test our new
procedure, described in Section \ref{sec:FOCM2L}, for estimating the
mass-to-light ratio efficiently. As input models we take both the
self-consistent isotropic rotator model described above and a model
constructed in the same way but including a dark matter halo. The halo
potential is of the form of equation (\ref{eqn:logpot}), with $r_0= 190$
arcsec ($9.7\kpc$), $v_0=220\;\kms$ and $q_\phi=1.0$. The
mass-to-light ratio of the stars in both input models is fixed to
$\Upsilon=5$.

\begin{figure}
\centering 
\includegraphics[angle=-90.0,width=0.95\hsize]{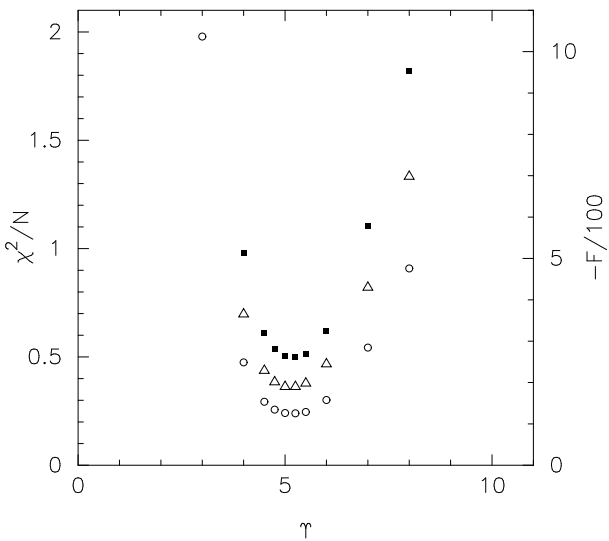}
\vskip0.2truecm
\caption[]{Quality of the particle model fit to the self-consistent
  isotropic rotator input model, as a function of assumed
  mass-to-light ratio $\Upsilon$.  $\chi^2$ values per data point are
  given for the particle model fit to the photometric and kinematic
  target observables (open circles) and to the kinematic observables
  alone (filled squares). The triangles correspond to the measured
  merit function $F$.  The input mass-to-light ratio is $\Upsilon=5$.
  All models are built using $\mu=100$.}
\label{fig:chi2m2l}
\end{figure}

\begin{figure}
\centering \includegraphics[angle=-90.0,width=0.85\hsize]{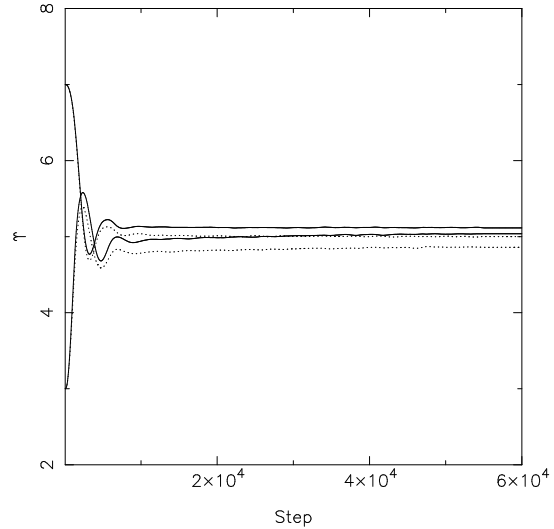}
\vskip0.2truecm
\caption[]{Direct mass-to-light ratio fits with NMAGIC. The plot shows
  the evolution of $\Upsilon$ with time during NMAGIC runs with
  different initial $\Upsilon$, for the self-consistent isotropic
  rotator target (solid lines), and the isotropic rotator in a
  potential including a dark halo (dotted lines).  The input
  mass-to-light ratio $\Upsilon=5$ in all cases. Time is given in
  terms of elapsed time steps where $10^4$ steps correspond to
  $\simeq40$ circular rotation periods at $1R_e$.}
\label{fig:m2levol}
\end{figure}

The results for a ``classical'' approach, in which we fit a dynamical
particle model to the data for different values of $\Upsilon$, are
presented in Figure \ref{fig:chi2m2l}, which shows the quality of the
fit as function of $\Upsilon$ for the self-consistent case. The input
value of $\Upsilon$ is recovered well. 
%
The results with the new procedure presented in Section
\ref{sec:FOCM2L} are summarized in Figure \ref{fig:m2levol}. The
figure shows the evolution of the mass-to-light ratio as a function of
time during NMAGIC model fits. Models for both the self-consistent
input galaxy and for the target model including a dark halo potential
are shown, with both low and high initial choices of $\Upsilon$.
%
The tests show that for the self-consistent case the input
mass-to-light ratio is recovered very well.  The uncertainties are
slightly larger when a dark matter halo is included, but the maximum
fractional error is less than three percent. We conclude that the new
scheme works very well and that $\Upsilon$ is recovered within a few
percent (for the amount and quality of data used in the present work).
The advantage of the new method is its efficiency, only one run is
needed to estimate $\Upsilon$ instead of order $10$, but at the cost
of not knowing the shape of $\chi^2$ as a function of $\Upsilon$ near
the minimum, i.e., the confidence interval.

\section{Dynamical models of NGC 4697}
\label{sec:models}
\begin{figure}
\centering 
\includegraphics[angle=0.0,width=0.85\hsize]{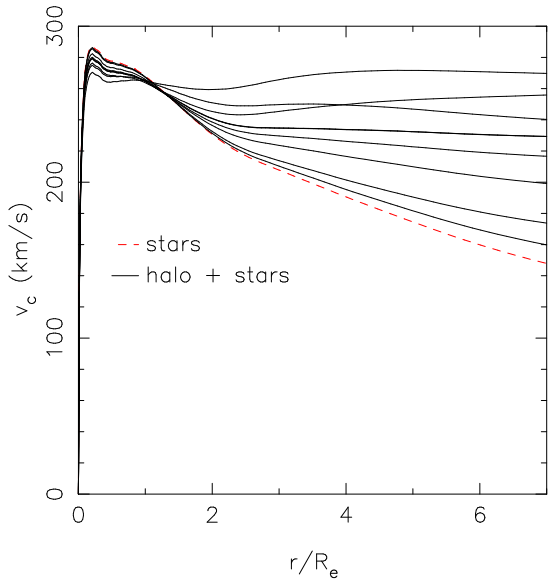}
\vskip0.2truecm
\caption[]{Circular velocity curves of the potentials used in the
  modelling, including the self-consistent model A (dashed line), and
  a sequence of dark matter halos (solid lines). The lines at
  $r/R_{e}=7$ run from model A (bottom) to K (top), with models F and
  G represented by the same curve; \cf Table \ref{tab:binmods}.}
\label{fig:circvel}
\end{figure}

After these tests we are now ready to use NMAGIC for constructing
axially symmetric dynamical models of NGC 4697. We investigate
self-consistent models as well as models including dark matter halos,
and fit the photometry, slit kinematics and PNe data. Our aim in this
paper is not to attempt to constrain detailed halo mass profiles, but
only to ascertain whether a dark matter component is allowed, or
required, by the kinematic data. Thus we investigate a simple sequence
of potentials A to K which include the contribution from the stellar
component and a halo potential as in equation~(\ref{eqn:logpot}), with
parameters given in Table \ref{tab:binmods}.  The parameters are
chosen to result in a sequence of circular speed curves ranging from
falling according to the distribution of stars to nearly flat over the
whole range of radii. This sequence is shown in Figure
\ref{fig:circvel}; all these circular velocity curves are computed in
the galaxy's equatorial plane and include the stellar component with
the respective best-fitting mass-to-light ratio as given in Table
\ref{tab:binmods}.
%

\begin{table*}
\vbox{\hfil
\begin{tabular}{cccccccccc}
  \hline
  \\[-1.7ex]
  \textsc{Halo} & $r_0/R_{e}$ & $v_0/\kms$ &
   $q_{\phi}$ & $\chi^2/N$ & $\chi^2_{alm}/N_{alm}$ &
   $\chi^2_{sl}/N_{sl}$ & $\chi^2_{PN}/N_{PN}$ & $-F$
   & $\Upsilon$ \\
  \\[-1.7ex] 
  \hline

  A &  $0$ & $0$   & $1.0$ & $0.453$ & $0.0323$ & $0.900$ & $0.968$ &
  $370.2$ & $5.78$ \\

  B &  $5.76$ & $80$  & $1.0$ & $0.415$ & $0.0254$ & $0.828$ & $0.884$ &
  $343.9$ & $5.74$ \\

  C &  $5.76$ & $120$ & $1.0$ & $0.439$ & $0.0343$ & $0.877$ & $0.784$ &
  $358.6$ & $5.71$ \\

  D &  $4.32$ & $160$ & $1.0$ & $0.404$ & $0.0288$ & $0.816$ & $0.610$ &
  $333.7$ & $5.58$ \\

  E &  $4.32$ & $190$ & $1.0$ & $0.404$ & $0.0244$ & $0.826$ & $0.520$ &
  $332.8$ & $5.49$ \\

  F &  $4.32$ & $210$ & $1.0$ & $0.386$ & $0.0229$ & $0.791$ & $0.476$ &
  $320.0$ & $5.45$ \\

  G &  $4.32$ & $210$ & $0.8$ & $0.382$ & $0.0203$ & $0.785$ & $0.439$ &
  $315.4$ & $5.46$ \\

  H &  $2.88$ & $210$ & $0.8$ & $0.376$ & $0.0232$ & $0.773$ & $0.397$ &
  $310.2$ & $5.28$ \\

  J &  $4.32$ & $250$ & $0.8$ & $0.383$ & $0.0242$ & $0.786$ & $0.377$ &
  $313.7$ & $5.34$ \\

  K &  $2.88$ & $250$ & $0.8$ & $0.377$ & $0.0212$ & $0.771$ & $0.506$ &
  $309.6$ & $5.10$ \\

  \hline
\end{tabular}
\hfil}
\caption{Table of model parameters and fit results. Columns (1)-(4) give the model 
  code and the parameters $r_0$, $v_0$ and $q_\phi$ used in equation (\ref{eqn:logpot})
  for the respective dark halo potential in this model. The next four columns list 
  the $\chi^2$ values per data point, for all observables [column (5)], and for the 
  density constraints, slit kinematic observables, and PN observables (data set
  PND1) separately [columns (6)-(8)]. Column (9) gives the numerical value of the 
  merit function in equation \ref{eqn:F}, and column (10) the final (r-band) 
  mass-to-light ratio. The respective number of constraints are $N=1316$, $N_{alm}=680$,
  $N_{sl}=604$, $N_{PN}=32$.}
\label{tab:binmods}
\end{table*}

\begin{table*}
\vbox{\hfil
\begin{tabular}{cccccccccc}
  \hline 
  \\[-1.7ex]  
  \textsc{Halo} & $r_0/R_{e}$ & $v_0/\kms$ & $q_{\phi}$ &
  $\chi^2/N$ & $\chi^2_{alm}/N_{alm}$ & $\chi^2_{sl}/N_{sl}$ &
  $-{\mathcal L}$ & $-F$ & $\Upsilon$ \\ 
  \\[-1.7ex]
  \hline

  A &  $0$ & $0$   & $1.0$ & $0.415$ & $0.0331$ & $0.845$ & $2042.9$ &   $2382.5$ & $5.81$ \\ 

  B &  $5.76$ & $80$  & $1.0$ & $0.405$ & $0.0282$ & $0.830$ & $2038.2$ & $2371.5$ & $5.76$ \\ 

  C &  $5.76$ & $120$ & $1.0$ & $0.419$ & $0.0331$ & $0.853$ & $2033.7$ & $2374.2$ & $5.72$ \\ 

  D &  $4.32$ & $160$ & $1.0$ & $0.406$ & $0.0314$ & $0.828$ & $2028.3$ & $2357.9$ & $5.60$ \\ 

  E &  $4.32$ & $190$ & $1.0$ & $0.391$ & $0.0271$ & $0.801$ & $2026.3$ & $2344.9$ & $5.54$ \\ 

  F &  $4.32$ & $210$ & $1.0$ & $0.402$ & $0.0304$ & $0.820$ & $2025.6$ & $2350.1$ & $5.49$ \\

  G &  $4.32$ & $210$ & $0.8$ & $0.396$ & $0.0232$ & $0.815$ & $2024.8$ & $2343.9$ & $5.48$ \\

  H &  $2.88$ & $210$ & $0.8$ & $0.373$ & $0.0245$ & $0.766$ & $2026.3$ & $2329.2$ & $5.31$ \\

  J &  $4.32$ & $250$ & $0.8$ & $0.374$ & $0.0203$ & $0.773$ & $2025.6$ & $2329.0$ & $5.37$ \\

  K &  $2.88$ & $250$ & $0.8$ & $0.369$ & $0.0198$ & $0.763$ & $2030.8$ & $2329.9$ & $5.14$ \\

  \hline
\end{tabular}
\hfil}
\caption{Table of model parameters and fit results,  similar to Table \ref{tab:binmods}, 
  but with all models computed using the likelihood scheme for the PNe as discrete
  kinematic tracers. Columns (8) and (9) now give the likelihood of the PN data set
  ${\mathcal L}$ and the merit function including ${\mathcal L}$ [equation \ref{eqn:FLH}].
  The other columns are equivalent to those in Table \ref{tab:binmods}.}
\label{tab:LH}
\end{table*}

To construct the models, we proceed as in Section \ref{sec:isorot}.
First, we compute the photometric observables. We expand the
deprojected luminosity distribution of NGC 4697 in a spherical
harmonics series on a grid of $40$ shells in radius,
quasi-logarithmically spaced with $r_{\rm min}=1.0\arcsec$ and $r_{\rm
  max}=700\arcsec$. As observables we use the luminosity on radial
shells $L_k$ and the higher order coefficients $A_{20,k}$, $A_{22,k}$,
$\cdots$, $A_{66,k}$ and $A_{80,k}$, at radii $r_k$. The $m\ne0$ terms
are set to zero to force the models to remain nearly axisymmetric,
i.e., within the limits set by the specified $A_{lm}$ errors. Because
the photometry is not seeing-deconvolved, for the innermost two points
($R<3"$) we only fit the $A_{00}$ term.  Errors for the luminosity
terms $A_{lm}$ are estimated by Monte Carlo simulations as in Section
\ref{sec:mu}. As kinematic constraints we use the luminosity weighted
Gauss-Hermite moments from the slit data, and the PNe kinematics,
either represented by binned line-of-sight velocity and velocity
dispersion points, or as discrete velocity measurements; see Sections
\ref{sec:pnevs} and \ref{sec:pnedisc}.

Again we fit particle models in a three step process. (i) First, we
start with the spherical particle model described in Section
\ref{sec:ics} and evolve it using NMAGIC to generate a self-consistent
particle realization with the luminosity distribution given by the
deprojection of the photometry.  (ii) Because NGC 4697 shows
significant rotation, we then switch retrograde particles similarly as
in Section \ref{sec:mu}, using $p_0=0.3$ and $L_0\simeq L_{\rm
  circ}(0.03 R_e)$. The resulting rotating particle model (hereafter,
model RIC) is used as a starting point to construct a series of
dynamical models by fitting the photometry and kinematics in different
halo potentials, as follows.  For every dark matter halo from
Table~\ref{tab:binmods}, we first relax RIC for $5000$ time steps in
the total gravitational potential, assuming a mass-to-light ratio of
5.74. For reference, $10000$ time steps in the self-consistent
potential correspond to $\approx 40$ circular rotation periods at
$R_{e}$ in spherical approximation. After this relaxation phase, we
evolve the particle system for $\sim 10^5$ NMAGIC correction steps
while fitting the complete set of constraints. During this correction
phase, the potential generated by the particles is updated after each
correction step but the dark matter potential (if present) is constant
in time. (iii) Subsequently, we keep the global potential constant and
evolve the system freely for another $5000$ steps, without further
correction steps. Models A, D, G and K were in addition evolved for a
further $10000$ steps with all potential terms active, to confirm that
the modest radial anisotropy required in these models does not lead to
dynamical instabilities. 

To make sure that the results are not biased by the way we incorporate
the PNe data, we have constructed three models in most halo
potentials. Each time the PNe data are represented differently, using
the binnings PND1, PND2, or the likelihood method.

The quality of the fit for different halo models can be characterized
by the quantity $F$ defined in equation (\ref{eqn:F}) or
(\ref{eqn:FLH}) and is given in Tables \ref{tab:binmods} and
\ref{tab:LH}. In addition, the value of $\chi^2$ per data point is
also shown, globally and for each data set separately, as are the
stellar mass-to-light ratios. For the same reasons as for the
isotropic rotator test models, the density constraints are very
accurately fit. The slit kinematics are typically fit within about
$0.9\sigma$ per point, slightly better than required. This is due to
the relatively low value used for the entropy smoothing, needed not to
bias the range of allowed potentials by the imposed smoothing.  
The PN $\chi^2$ values indicate that the PN data are consistent with all
models.

The likelihood values for the models reported in Table \ref{tab:LH}
would formally allow us to exclude a large fraction of the halo
potentials tested, using $\Delta{\mathcal L}=0.5$.  However, these
likelihood differences depend to a significant degree on very few PNe
in the wings of the LOSVDs.  To assess this we have compared
histograms of PN velocities with the LOSVDs of these models in cones
along the major and minor axes. A $\chi^2$ test shows that all model
LOSVDs except for the model without dark halo are formally consistent
with the PNe velocity histograms and associated Poisson errors.
However, the low dark matter models show systematic deviations in form
from the more flat-topped data histograms, which we judge significant
for models A to C.  Also, for these models the LOSVD $\chi^2$ and the
likelihood are correlated, which is not the case for the more massive
halo models.  On the basis of these findings we believe the likelihood
results can probably be trusted for ruling out models A-C, but not for
any of the more massive halo models. Thus we estimate that the range
of circular velocities at $5R_e$ consistent with the data is
$v_c(5R_e)\gta 200\kms$, with the best models having $v_c(5R_e)\simeq
250\kms$.  To reach more stringent constraints will require PN
velocities at even larger radii. 

\begin{figure}
\centering 
\includegraphics[angle=-90.0,width=0.95\hsize]{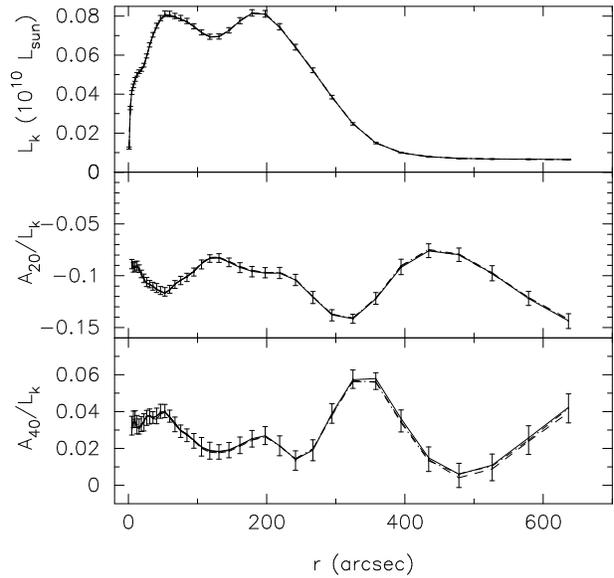}
\vskip0.2truecm
\caption[]{Comparison of the photometric constraints with the final
  models A (self-consistent, dashed), D (dotted), and G (full line).
  The points correspond to the target input data. From top to bottom:
  luminosity on radial shells profile $L_k=\sqrt{4\pi} A_{00}$, and
  normalized $A_{20}$ and $A_{40}$ profiles.}
\label{fig:almfit}
\end{figure}

\begin{figure*}
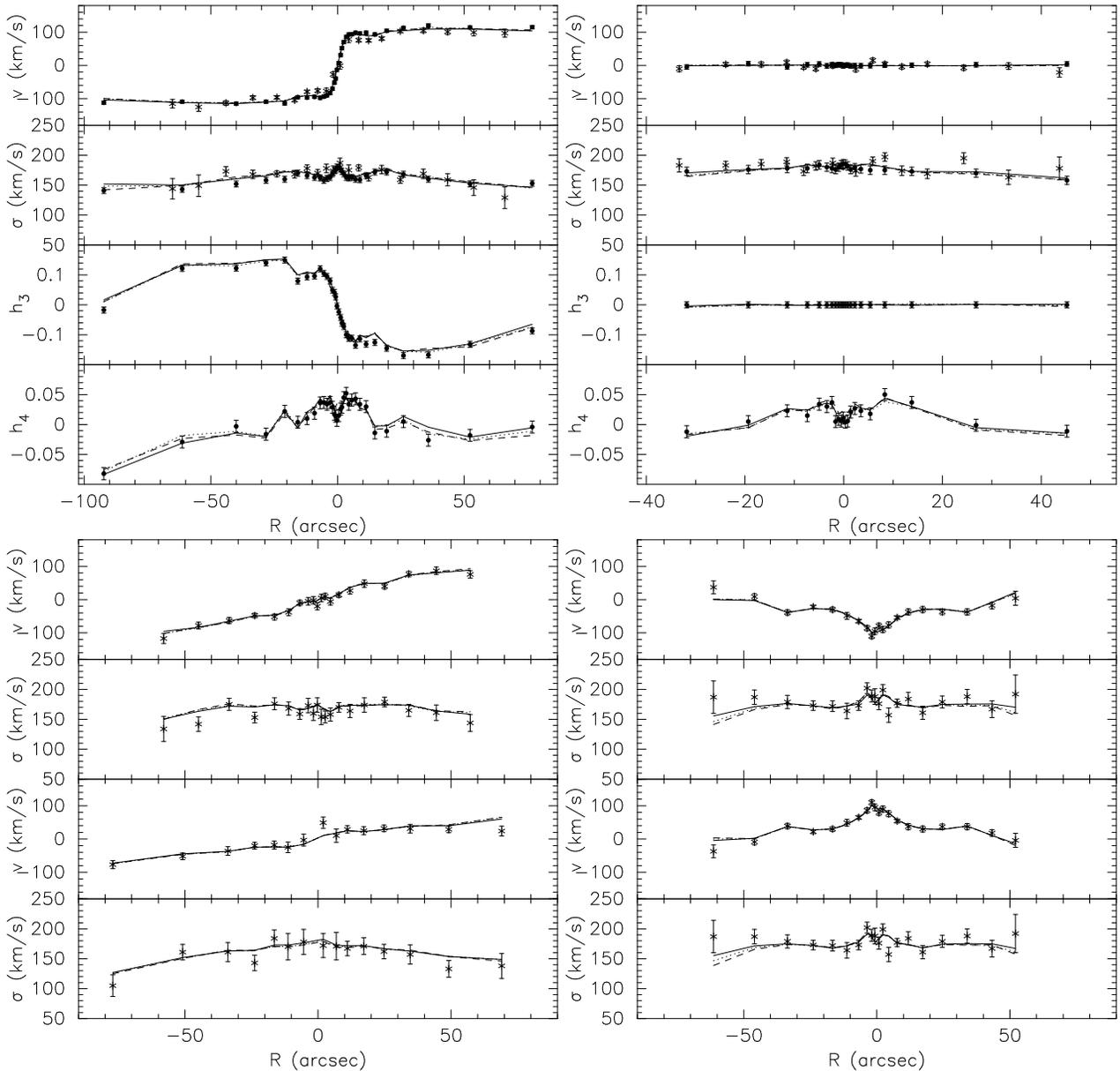

\includegraphics[angle=-90.0,width=0.47\hsize]{fig17a_mjfit.ps}
\includegraphics[angle=-90.0,width=0.47\hsize]{fig17b_mnfit.ps}
\includegraphics[angle=-90.0,width=0.47\hsize]{fig17c_parallmjfit.ps}
\includegraphics[angle=-90.0,width=0.47\hsize]{fig17d_parallmnfit.ps}
\vskip0.2truecm
\vskip0.2truecm
\caption[]{Comparison of models A, D, and G to the absorption line
  kinematic data along the major axis (top left), minor axis (top
  right), the slits parallel to the major axis (bottom left), and the
  slits parallel to the minor axis (bottom right). Full and starred
  data points show our new data and the BDI data, respectively. The
  model data points are averages over the same slit cells as the
  target data (see Fig.~\ref{fig:slits}), and are connected by
  straight line segments. Linestyles for the models are the same as in
  Fig.~\ref{fig:almfit}.}
\label{fig:alsfit}
\end{figure*}

\begin{figure}
\centering 
\includegraphics[angle=-90.0,width=0.95\hsize]{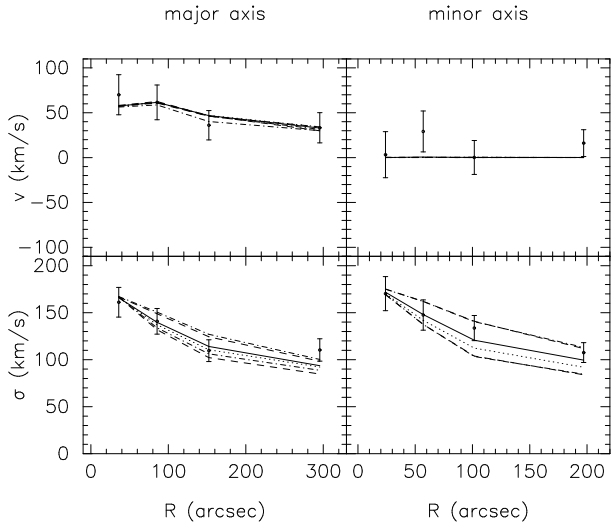}
\vskip0.2truecm
\caption[]{Comparison of the PNe velocity and velocity dispersion data
  (PND1, points) with models A, D, G, and K. Top left: $v$ along the
  positive major axis. Top right: The same for the minor axis. Bottom
  left: $\sigma$ along the positive major axis. Bottom right: The same
  but for the minor axis. Dashed, dotted, full, and upper dashed lines
  show models A, D, G, and K; the two dash-dotted lines show the
  variants of models A and K obtained with the likelihood scheme for
  the PNe.}
\label{fig:pnefit}
\end{figure}

Figures \ref{fig:almfit}, \ref{fig:alsfit}, \ref{fig:pnefit} present
results from some of these models, comparing the stars-only model A
and the three halo models D, G, and K to the data.  Figure
\ref{fig:almfit} shows the comparison of models A, D, G with the
photometric constraints.  The model lines match the target data points
perfectly, in accordance with the very small $\chi^2_{alm}/N_{alm}$
values in Table \ref{tab:binmods}.  Figure \ref{fig:alsfit} compares
the projected absorption line kinematics of the three models with our
measurements and the BDI data.  The fits are generally excellent.
Along the major and minor axes one can see how the models have found
compromises to deal with asymmetries of the data on both sides of the
galaxy, and slight discrepancies between our and the BDI data, e.g.,
in the region around $\pm 10 \arcsec$ along the major axis. As for the
isotropic rotator, the major axis higher order moments in
Fig.~\ref{fig:alsfit} are even somewhat overfitted; see the discussion
in Section~\ref{sec:mu}.

Figure \ref{fig:pnefit} compares the final A, D, G, and K models with
the PNe kinematic constraints along the major axis (left) and minor
axis (right); on each axis we show mean velocity (top) and velocity
dispersion (bottom). The model curves in Fig.~\ref{fig:pnefit} and the
$\chi^2$ per data point values in Table \ref{tab:binmods} are computed
for PN dataset PND1. There was no difference between these values and
those obtained with PN dataset PND2 in all cases where we modelled
both. The two additional lines in the panels of Fig.~\ref{fig:pnefit}
show the mean velocities and velocity dispersions for the variants of
models A, K obtained with the likelihood scheme for the PNe (see
Section \ref{sec:pnedisc}), computed by binning the particles in these
models a posteriori in the same bins as for dataset PND1.  While there
is little difference for model K, the likelihood variant of model A
fits the observed PN data points actually better than the original
model A based on the PND1 data.

%
%
%
\begin{figure}
\centering 
\includegraphics[angle=-90.0,width=0.95\hsize]{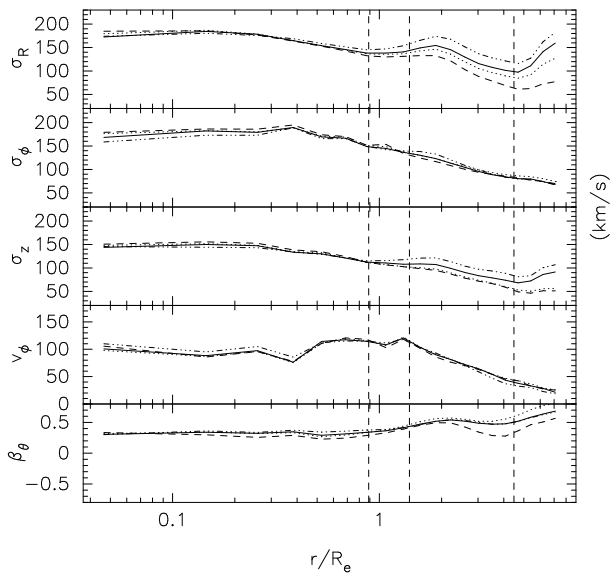}
\vskip0.2truecm
\caption[]{Internal velocity moments in the equatorial plane for
  models A, D, G, K (dashed, dotted, full, and dash-triple dotted
  lines, respectively). The vertical dashed lines indicate the radial
  extent of the minor axis slit data, major axis slit data, and PN
  data, from left to right.}
\label{fig:intkin4697_mj}
\end{figure}

Overall, this figure illustrates that with increasing halo mass the
fit to the PNe kinematic data improves slightly.  Models G and K
bracket the best-fitting models to the binned data in
Fig.~\ref{fig:pnefit}.  However, also model A without dark matter
still has a $\chi^2_{PN}/N_{PN}$ just below one, despite being
systematically a little low in the minor axis dispersion plot. When we
force the self-consistent model to improve the PN data fit at the
expense of the slit kinematic data fit, the model starts to develop
specific anisotropy features at the radii of the outer PN data. This
suggests that with PN data extending to somewhat larger radii,
$\approx 400$ arcsec ($R\simeq6R_e)$, the model without dark halo
might start to fail  also for the binned data in
Fig.~\ref{fig:pnefit}, consistent with the likelihood and LOSVD
results.  In conclusion, a variety of dark halos with $v_c(5R_e)\gta
200\kms$ are consistent with all the kinematic data currently
available for NGC 4697.

Finally, Figure \ref{fig:intkin4697_mj} shows the internal kinematics
of the particle models A, D, G, and K. The upper panels give $\sigma_R$,
$\sigma_\phi$ and $\sigma_z$, followed by $v_\phi$. The last panel
displays the anisotropy parameter
$\beta_\theta=1-\sigma^2_\theta/\sigma^2_r$, which is zero for an
isotropic rotator model. All quantities are given as averages over the
models' equatorial plane.  The more massive halo models become more
radially anisotropic in the outer parts in terms of $\sigma_R$ vs.\
$\sigma_\phi$, but $\beta_\theta$ does not increase beyond model D
because $\sigma_R$ and $\sigma_z$ increase in parallel while
$\sigma_\phi$ remains constant.  Thus the additional kinetic energy
that stars at large radii must have in these models, is hidden
in the plane of the sky.  Conversely, at small radii the velocity
dispersions in models G-K are slightly lower, compensating for the
larger radial velocities of halo stars along the line-of-sight to the
center. These models have $\beta\simeq0.3$ at the center, which
increases with radius and reaches $\beta\simeq0.5$ at $\gta2 R_{e}$.

\section{Summary and Conclusions}
\label{sec:conclusions}
In this paper, we have presented new surface brightness measurements
and long slit spectroscopic data for the E4 galaxy NGC 4697, and
combined these data with existing long slit kinematics and discrete
PNe position and velocity measurements to construct dynamical models
for this galaxy. The combined data set runs from the center of the
galaxy to about 4.5 effective radii.

For the first time, we have modelled such a dataset with the new and
flexible $\chi^2$-made-to-measure ($\chi^2$M2M) particle code NMAGIC.
We have extended NMAGIC to include seeing effects and have implemented
an efficient method to estimate the mass-to-light ratio $\Upsilon$.
Tests of this scheme using isotropic rotator input models have shown
that the method recovers $\Upsilon$ within a few percent both for
self-consistent and dark matter dominated target galaxies.  In
addition, we have implemented a likelihood scheme which allows us to
treat the PNe as discrete velocity measurements, so that no binning in
velocities is needed. The modelling presented in this paper shows that
the $\chi^2$M2M/NMAGIC particle method is now competitive with the
familiar Schwarzschild method. In fact, it has already gone further in
that the gravitational potential of the stars has been allowed to vary
in the modelling, the mass-to-light ratio has been adapted on the fly,
the stability of the models has been checked, and, in \citet{delo+07},
NMAGIC has been used to construct triaxial and rotating triaxial
models.

Even though NMAGIC does not require any symmetry assumptions for the
modelling, we have in this paper forced the method to generate
axisymmetric particle models for NGC 4697. Both self-consistent models
without dark matter, and models following a sequence of circular speed
curves with increasing dark halo contributions have been investigated.
The PN data have been used both binned on two different spatial grids,
as well as with the new likelihood scheme, to make sure that the
results are not biased by the way the PNe data are incorporated.

Our main astronomical result is that models with a variety
of dark matter halos are consistent with all the data for NGC 4697, as
long as the circular velocity $v_c(5R_e) \gta 200\kms$ at $5R_e$.
These models include potentials with sufficiently massive halos to
generate nearly flat circular rotation curves.  These models fit all
kinematic data with $\chi^2/N<1$, including the PN LOSVDs.  Models
with no dark matter are not consistent with the PN LOSVDs, as judged
from both the LOSVD histograms and their likelihood values.  Amongst
the acceptable models, the more massive dark halo models with
$v_c(5R_e)\simeq250\kms$ tend to fit the data slightly better in the
sense of lower $\chi^2/N$, for both the slit kinematics and the PN
data, but these variations are small and not yet statistically
significant. To further narrow down the range of acceptable dark
matter models would require PN velocities at even larger radii than
currently available, out to an estimated $\simeq 6 R_e$ from the
center.

Our models differ from earlier studies performed by
\citet{mendez_etal01} in the sense that we generate axisymmetric
models instead of spherical ones and that our models are flexible with
regard to anisotropy.  The best-fitting models are slightly radially
anisotropic, with $\beta\simeq0.3$ at the center, increasing to
$\beta\simeq0.5$ at $\gta2 R_{e}$. This is consistent with the value given
by \citet{dekel+05} obtained from merger simulations carried out
within the $\Lambda$CDM cosmology framework.

\bigskip
\noindent

\section*{Acknowledgments}
We thank Lodovico Coccato for help with the data and the anonymous
referee for comments that led to a better understanding of the effect
of entropy on the models' smoothness. VPD is grateful for a Brooks
Prize Fellowship while at the University of Washington. RHM would like
to acknowledge support by the U.S. National Science Foundation under
Grant 0307489.

\bibliography{mybib}

\begin{thebibliography}{}

\bibitem[\protect\citeauthoryear{{Arnaboldi}, {Freeman}, {Gerhard}, {Matthias},
  {Kudritzki}, {M{\'e}ndez}, {Capaccioli} \& {Ford}}{{Arnaboldi}
  et~al.}{1998}]{arnaboldi+98}
{Arnaboldi} M.,  {Freeman} K.~C.,  {Gerhard} O.,  {Matthias} M.,  {Kudritzki}
  R.~P.,  {M{\'e}ndez} R.~H.,  {Capaccioli} M.,    {Ford} H.,  1998, \apj, 507,
  759

\bibitem[\protect\citeauthoryear{{Arnaboldi}, {Freeman}, {Mendez},
  {Capaccioli}, {Ciardullo}, {Ford}, {Gerhard}, {Hui}, {Jacoby}, {Kudritzki} \&
  {Quinn}}{{Arnaboldi} et~al.}{1996}]{arnaboldi+96}
{Arnaboldi} M.,  {Freeman} K.~C.,  {Mendez} R.~H.,  {Capaccioli} M.,
  {Ciardullo} R.,  {Ford} H.,  {Gerhard} O.,  {Hui} X.,  {Jacoby} G.~H.,
  {Kudritzki} R.~P.,    {Quinn} P.~J.,  1996, \apj, 472, 145

\bibitem[\protect\citeauthoryear{{Awaki}, {Mushotzky}, {Tsuru}, {Fabian},
  {Fukazawa}, {Loewenstein}, {Makishima}, {Matsumoto}, {Matsushita}, {Mihara},
  {Ohashi}, {Ricker}, {Serlemitsos}, {Tsusaka} \& {Yamazaki}}{{Awaki}
  et~al.}{1994}]{awaki+94}
{Awaki} H.,  {Mushotzky} R.,  {Tsuru} T.,  {Fabian} A.~C.,  {Fukazawa} Y.,
  {Loewenstein} M.,  {Makishima} K.,  {Matsumoto} H.,  {Matsushita} K.,
  {Mihara} T.,  {Ohashi} T.,  {Ricker} G.~R.,  {Serlemitsos} P.~J.,  {Tsusaka}
  Y.,    {Yamazaki} T.,  1994, \pasj, 46, L65

\bibitem[\protect\citeauthoryear{{Bender} \& {Moellenhoff}}{{Bender} \&
  {Moellenhoff}}{1987}]{bender_moell87}
{Bender} R.,  {Moellenhoff} C.,  1987, \aap, 177, 71

\bibitem[\protect\citeauthoryear{{Bender}, {Saglia} \& {Gerhard}}{{Bender}
  et~al.}{1994}]{BSG94}
{Bender} R.,  {Saglia} R.,    {Gerhard} O.,  1994, \mnras, 269, 785

\bibitem[\protect\citeauthoryear{{Binney} \& {Tremaine}}{{Binney} \&
  {Tremaine}}{1987}]{bin_tre87}
{Binney} J.,  {Tremaine} S.,  1987, {Galactic Dynamics}.
Princeton, NJ, Princeton University Press

\bibitem[\protect\citeauthoryear{{Binney}, {Davies} \& {Illingworth}}{{Binney}
  et~al.}{1990}]{binney_etal90}
{Binney} J.~J.,  {Davies} R.~L.,    {Illingworth} G.~D.,  1990, \apj, 361, 78

\bibitem[\protect\citeauthoryear{{Cappellari}, {Bacon}, {Bureau}, {Damen},
  {Davies}, {de Zeeuw}, {Emsellem}, {Falc{\'o}n-Barroso}, {Krajnovi{\'c}},
  {Kuntschner}, {McDermid}, {Peletier}, {Sarzi}, {van den Bosch} \& {van de
  Ven}}{{Cappellari} et~al.}{2006}]{cappellari+06}
{Cappellari} M.,  {Bacon} R.,  {Bureau} M.,  {Damen} M.~C.,  {Davies} R.~L.,
  {de Zeeuw} P.~T.,  {Emsellem} E.,  {Falc{\'o}n-Barroso} J.,  {Krajnovi{\'c}}
  D.,  {Kuntschner} H.,  {McDermid} R.~M.,  {Peletier} R.~F.,  {Sarzi} M.,
  {van den Bosch} R.~C.~E.,    {van de Ven} G.,  2006, \mnras, 366, 1126

\bibitem[\protect\citeauthoryear{{Carollo}, {de Zeeuw} \& {van der
  Marel}}{{Carollo} et~al.}{1995}]{carollo_etal95}
{Carollo} C.~M.,  {de Zeeuw} P.~T.,    {van der Marel} R.~P.,  1995, \mnras,
  276, 1131

\bibitem[\protect\citeauthoryear{{Carter}}{{Carter}}{1987}]{carter87}
{Carter} D.,  1987, \apj, 312, 514

\bibitem[\protect\citeauthoryear{{de Blok}, {Bosma} \& {McGaugh}}{{de Blok}
  et~al.}{2003}]{deblok+03}
{de Blok} W.~J.~G.,  {Bosma} A.,    {McGaugh} S.,  2003, \mnras, 340, 657

\bibitem[\protect\citeauthoryear{{de Lorenzi}, {Debattista}, {Gerhard} \&
  {Sambhus}}{{de Lorenzi} et~al.}{2007}]{delo+07}
{de Lorenzi} F.,  {Debattista} V.~P.,  {Gerhard} O.,    {Sambhus} N.,  2007,
  \mnras, 376, 71

\bibitem[\protect\citeauthoryear{{Debattista} \& {Sellwood}}{{Debattista} \&
  {Sellwood}}{2000}]{victor_sell00}
{Debattista} V.~P.,  {Sellwood} J.~A.,  2000, \apj, 543, 704

\bibitem[\protect\citeauthoryear{{Dehnen}}{{Dehnen}}{1993}]{dehnen93}
{Dehnen} W.,  1993, \mnras, 265, 250

\bibitem[\protect\citeauthoryear{{Dehnen} \& {Gerhard}}{{Dehnen} \&
  {Gerhard}}{1994}]{dehnen_ger94}
{Dehnen} W.,  {Gerhard} O.~E.,  1994, \mnras, 268, 1019

\bibitem[\protect\citeauthoryear{{Dejonghe}, {de Bruyne}, {Vauterin} \&
  {Zeilinger}}{{Dejonghe} et~al.}{1996}]{dejonghe_etal96}
{Dejonghe} H.,  {de Bruyne} V.,  {Vauterin} P.,    {Zeilinger} W.~W.,  1996,
  \aap, 306, 363

\bibitem[\protect\citeauthoryear{{Dekel}, {Stoehr}, {Mamon}, {Cox}, {Novak} \&
  {Primack}}{{Dekel} et~al.}{2005}]{dekel+05}
{Dekel} A.,  {Stoehr} F.,  {Mamon} G.~A.,  {Cox} T.~J.,  {Novak} G.~S.,
  {Primack} J.~R.,  2005, \nat, 437, 707

\bibitem[\protect\citeauthoryear{{Douglas}, {Napolitano}, {Romanowsky},
  {Coccato}, {Kuijken}, {Merrifield}, {Arnaboldi}, {Gerhard}, {Freeman},
  {Merrett}, {Noordermeer} \& {Capaccioli}}{{Douglas}
  et~al.}{2007}]{douglas+07}
{Douglas} N.~G.,  {Napolitano} N.~R.,  {Romanowsky} A.~J.,  {Coccato} L.,
  {Kuijken} K.,  {Merrifield} M.~R.,  {Arnaboldi} M.,  {Gerhard} O.,  {Freeman}
  K.~C.,  {Merrett} H.~R.,  {Noordermeer} E.,    {Capaccioli} M.,  2007, \apj,
  664, 257

\bibitem[\protect\citeauthoryear{{Fried} \& {Illingworth}}{{Fried} \&
  {Illingworth}}{1994}]{FI94}
{Fried} J.,  {Illingworth} G.,  1994, \aj, 107, 992

\bibitem[\protect\citeauthoryear{{Gebhardt}, {Richstone}, {Tremaine}, {Lauer},
  {Bender}, {Bower}, {Dressler}, {Faber}, {Filippenko}, {Green}, {Grillmair},
  {Ho}, {Kormendy}, {Magorrian} \& {Pinkney}}{{Gebhardt}
  et~al.}{2003}]{gebhardt_etal03}
{Gebhardt} K.,  {Richstone} D.,  {Tremaine} S.,  {Lauer} T.~R.,  {Bender} R.,
  {Bower} G.,  {Dressler} A.,  {Faber} S.~M.,  {Filippenko} A.~V.,  {Green} R.,
   {Grillmair} C.,  {Ho} L.~C.,  {Kormendy} J.,  {Magorrian} J.,    {Pinkney}
  J.,  2003, \apj, 583, 92

\bibitem[\protect\citeauthoryear{{Gerhard}, {Jeske}, {Saglia} \&
  {Bender}}{{Gerhard} et~al.}{1998}]{gerhard_etal98}
{Gerhard} O.,  {Jeske} G.,  {Saglia} R.~P.,    {Bender} R.,  1998, \mnras, 295,
  197

\bibitem[\protect\citeauthoryear{{Gerhard}, {Kronawitter}, {Saglia} \&
  {Bender}}{{Gerhard} et~al.}{2001}]{gerhard+01}
{Gerhard} O.,  {Kronawitter} A.,  {Saglia} R.~P.,    {Bender} R.,  2001, \aj,
  121, 1936

\bibitem[\protect\citeauthoryear{{Gerhard}}{{Gerhard}}{1993}]{gerhard93}
{Gerhard} O.~E.,  1993, \mnras, 265, 213

\bibitem[\protect\citeauthoryear{{Gerhard} \& {Binney}}{{Gerhard} \&
  {Binney}}{1996}]{gerhard_bin96}
{Gerhard} O.~E.,  {Binney} J.~J.,  1996, \mnras, 279, 993

\bibitem[\protect\citeauthoryear{{Goudfrooij}, {Hansen}, {Jorgensen},
  {Norgaard-Nielsen}, {de Jong} \& {van den Hoek}}{{Goudfrooij}
  et~al.}{1994}]{goudfrooij_etal94}
{Goudfrooij} P.,  {Hansen} L.,  {Jorgensen} H.~E.,  {Norgaard-Nielsen} H.~U.,
  {de Jong} T.,    {van den Hoek} L.~B.,  1994, \aaps, 104, 179

\bibitem[\protect\citeauthoryear{{Griffiths}, {Casertano}, {Im} \&
  {Ratnatunga}}{{Griffiths} et~al.}{1996}]{griffiths+96}
{Griffiths} R.~E.,  {Casertano} S.,  {Im} M.,    {Ratnatunga} K.~U.,  1996,
  \mnras, 282, 1159

\bibitem[\protect\citeauthoryear{{Hernquist}}{{Hernquist}}{1990}]{hernquist90}
{Hernquist} L.,  1990, \apj, 356, 359

\bibitem[\protect\citeauthoryear{{Hui}, {Ford}, {Freeman} \& {Dopita}}{{Hui}
  et~al.}{1995}]{hui_etal95}
{Hui} X.,  {Ford} H.~C.,  {Freeman} K.~C.,    {Dopita} M.~A.,  1995, \apj, 449,
  592

\bibitem[\protect\citeauthoryear{{Humphrey}, {Buote}, {Gastaldello},
  {Zappacosta}, {Bullock}, {Brighenti} \& {Mathews}}{{Humphrey}
  et~al.}{2006}]{humphrey+06}
{Humphrey} P.~J.,  {Buote} D.~A.,  {Gastaldello} F.,  {Zappacosta} L.,
  {Bullock} J.~S.,  {Brighenti} F.,    {Mathews} W.~G.,  2006, \apj, 646, 899

\bibitem[\protect\citeauthoryear{{Illingworth} \& {Schechter}}{{Illingworth} \&
  {Schechter}}{1982}]{IS82}
{Illingworth} G.~D.,  {Schechter} P.~L.,  1982, \apj, 256, 481

\bibitem[\protect\citeauthoryear{{Irwin}, {Sarazin} \& {Bregman}}{{Irwin}
  et~al.}{2000}]{irwin_etal00}
{Irwin} J.~A.,  {Sarazin} C.~L.,    {Bregman} J.~N.,  2000, \apj, 544, 293

\bibitem[\protect\citeauthoryear{{Kalnajs}}{{Kalnajs}}{1977}]{kalnajs77}
{Kalnajs} A.~J.,  1977, \apj, 212, 637

\bibitem[\protect\citeauthoryear{{Kochanek} \& {Rybicki}}{{Kochanek} \&
  {Rybicki}}{1996}]{kochanek_ryb96}
{Kochanek} C.~S.,  {Rybicki} G.~B.,  1996, \mnras, 280, 1257

\bibitem[\protect\citeauthoryear{{Koprolin}, A. \& {Zeilinger}}{{Koprolin}
  et~al.}{2000}]{KZ00}
{Koprolin} W.,  A.   {Zeilinger} W.,  2000, \aaps, 145, 71

\bibitem[\protect\citeauthoryear{{Kronawitter}, {Saglia}, {Gerhard} \&
  {Bender}}{{Kronawitter} et~al.}{2000}]{krona_etal00}
{Kronawitter} A.,  {Saglia} R.~P.,  {Gerhard} O.,    {Bender} R.,  2000, \aaps,
  144, 53

\bibitem[\protect\citeauthoryear{{Loewenstein} \& {White} III}{{Loewenstein} \&
  {White}}{1999}]{loewenstein+99}
{Loewenstein} M.,  {White} III R.~E.,  1999, \apj, 518, 50

\bibitem[\protect\citeauthoryear{{Lynden-Bell}}{{Lynden-Bell}}{1962}]{lyndenbe%
ll62}
{Lynden-Bell} D.,  1962, \mnras, 123, 447

\bibitem[\protect\citeauthoryear{{Magorrian}}{{Magorrian}}{1999}]{magorrian99}
{Magorrian} J.,  1999, \mnras, 302, 530

\bibitem[\protect\citeauthoryear{{Magorrian} \& {Binney}}{{Magorrian} \&
  {Binney}}{1994}]{magorrian_bin94}
{Magorrian} J.,  {Binney} J.,  1994, \mnras, 271, 949

\bibitem[\protect\citeauthoryear{{McGaugh}, {de Blok}, {Schombert}, {Kuzio de
  Naray} \& {Kim}}{{McGaugh} et~al.}{2007}]{mcgaugh+07}
{McGaugh} S.~S.,  {de Blok} W.~J.~G.,  {Schombert} J.~M.,  {Kuzio de Naray} R.,
     {Kim} J.~H.,  2007, \apj, 659, 149

\bibitem[\protect\citeauthoryear{{Mehlert}, {Saglia}, {Bender} \&
  {Wegner}}{{Mehlert} et~al.}{2000}]{Mehlert+00}
{Mehlert} D.,  {Saglia} R.,  {Bender} R.,    {Wegner} G.,  2000, \aaps, 141,
  449

\bibitem[\protect\citeauthoryear{{M{\'e}ndez}, {Riffeser}, {Kudritzki},
  {Matthias}, {Freeman}, {Arnaboldi}, {Capaccioli} \& {Gerhard}}{{M{\'e}ndez}
  et~al.}{2001}]{mendez_etal01}
{M{\'e}ndez} R.~H.,  {Riffeser} A.,  {Kudritzki} R.-P.,  {Matthias} M.,
  {Freeman} K.~C.,  {Arnaboldi} M.,  {Capaccioli} M.,    {Gerhard} O.~E.,
  2001, \apj, 563, 135

\bibitem[\protect\citeauthoryear{{M{\'e}ndez}, {Thomas}, {Saglia}, {Maraston},
  {Kudritzki} \& {Bender}}{{M{\'e}ndez} et~al.}{2005}]{mendez+05}
{M{\'e}ndez} R.~H.,  {Thomas} D.,  {Saglia} R.~P.,  {Maraston} C.,  {Kudritzki}
  R.~P.,    {Bender} R.,  2005, \apj, 627, 767

\bibitem[\protect\citeauthoryear{{Moore}, {Quinn}, {Governato}, {Stadel} \&
  {Lake}}{{Moore} et~al.}{1999}]{moore+99}
{Moore} B.,  {Quinn} T.,  {Governato} F.,  {Stadel} J.,    {Lake} G.,  1999,
  \mnras, 310, 1147

\bibitem[\protect\citeauthoryear{{Navarro}, {Frenk} \& {White}}{{Navarro}
  et~al.}{1996}]{navarro+96}
{Navarro} J.~F.,  {Frenk} C.~S.,    {White} S.~D.~M.,  1996, \apj, 462, 563

\bibitem[\protect\citeauthoryear{{Peletier}, {Davies}, {Illingworth}, {Davis}
  \& {Cawson}}{{Peletier} et~al.}{1990}]{Peletier1990}
{Peletier} R.~F.,  {Davies} R.~L.,  {Illingworth} G.~D.,  {Davis} L.~E.,
  {Cawson} M.,  1990, \aj, 100, 1091

\bibitem[\protect\citeauthoryear{{Pinkney}, {Gebhardt}, {Bender}, {Bower},
  {Dressler}, {Faber}, {Filippenko}, {Green}, {Ho}, {Kormendy}, {Lauer},
  {Magorrian}, {Richstone} \& {Tremaine}}{{Pinkney}
  et~al.}{2003}]{pinkney_etal03}
{Pinkney} J.,  {Gebhardt} K.,  {Bender} R.,  {Bower} G.,  {Dressler} A.,
  {Faber} S.~M.,  {Filippenko} A.~V.,  {Green} R.,  {Ho} L.~C.,  {Kormendy} J.,
   {Lauer} T.~R.,  {Magorrian} J.,  {Richstone} D.,    {Tremaine} S.,  2003,
  \apj, 596, 903

\bibitem[\protect\citeauthoryear{{Pinkney}, {Gebhardt}, {Richstone} \& {Nuker
  Team}}{{Pinkney} et~al.}{2000}]{pinkney_etal00}
{Pinkney} J.,  {Gebhardt} K.,  {Richstone} D.,    {Nuker Team} 2000, Bulletin
  of the American Astronomical Society, 32, 1437

\bibitem[\protect\citeauthoryear{{Rix}, {de Zeeuw}, {Cretton}, {van der Marel}
  \& {Carollo}}{{Rix} et~al.}{1997}]{rix_etal97}
{Rix} H.-W.,  {de Zeeuw} P.~T.,  {Cretton} N.,  {van der Marel} R.~P.,
  {Carollo} C.~M.,  1997, \apj, 488, 702

\bibitem[\protect\citeauthoryear{{Romanowsky}, {Douglas}, {Arnaboldi},
  {Kuijken}, {Merrifield}, {Napolitano}, {Capaccioli} \&
  {Freeman}}{{Romanowsky} et~al.}{2003}]{romanowsky_etal03}
{Romanowsky} A.~J.,  {Douglas} N.~G.,  {Arnaboldi} M.,  {Kuijken} K.,
  {Merrifield} M.~R.,  {Napolitano} N.~R.,  {Capaccioli} M.,    {Freeman}
  K.~C.,  2003, Science, 301, 1696

\bibitem[\protect\citeauthoryear{{Romanowsky} \& {Kochanek}}{{Romanowsky} \&
  {Kochanek}}{2001}]{romano_koch01}
{Romanowsky} A.~J.,  {Kochanek} C.~S.,  2001, \apj, 553, 722

\bibitem[\protect\citeauthoryear{{Rusin} \& {Kochanek}}{{Rusin} \&
  {Kochanek}}{2005}]{rusin+kochanek05}
{Rusin} D.,  {Kochanek} C.~S.,  2005, \apj, 623, 666

\bibitem[\protect\citeauthoryear{{Rybicki}}{{Rybicki}}{1987}]{rybicki87}
{Rybicki} G.~B.,  1987, in {de Zeeuw} P.~T.,  ed., {IAU Symp. 127: Structure
  and Dynamics of Elliptical Galaxies} p.~397

\bibitem[\protect\citeauthoryear{{Sambhus}, {Gerhard} \&
  {M{\'e}ndez}}{{Sambhus} et~al.}{2006}]{niri_etal06}
{Sambhus} N.,  {Gerhard} O.,    {M{\'e}ndez} R.~H.,  2006, \aj, 131, 837

\bibitem[\protect\citeauthoryear{{Sansom}, {Hibbard} \& {Schweizer}}{{Sansom}
  et~al.}{2000}]{sansom_etal00}
{Sansom} A.~E.,  {Hibbard} J.~E.,    {Schweizer} F.,  2000, \aj, 120, 1946

\bibitem[\protect\citeauthoryear{{Scorza} \& {Bender}}{{Scorza} \&
  {Bender}}{1995}]{scorza+95}
{Scorza} C.,  {Bender} R.,  1995, \aap, 293, 20

\bibitem[\protect\citeauthoryear{{Scorza}, {Bender}, {Winkelmann}, {Capaccioli}
  \& {Macchetto}}{{Scorza} et~al.}{1998}]{scorza+98}
{Scorza} C.,  {Bender} R.,  {Winkelmann} C.,  {Capaccioli} M.,    {Macchetto}
  D.~F.,  1998, \aaps, 131, 265

\bibitem[\protect\citeauthoryear{{Sellwood}}{{Sellwood}}{2003}]{sellwood03}
{Sellwood} J.~A.,  2003, \apj, 587, 638

\bibitem[\protect\citeauthoryear{{Syer} \& {Tremaine}}{{Syer} \&
  {Tremaine}}{1996}]{ST96}
{Syer} D.,  {Tremaine} S.,  1996, \mnras, 282, 223

\bibitem[\protect\citeauthoryear{{Thomas}, {Saglia}, {Bender}, {Thomas},
  {Gebhardt}, {Magorrian}, {Corsini} \& {Wegner}}{{Thomas}
  et~al.}{2005}]{thomas_etal05}
{Thomas} J.,  {Saglia} R.~P.,  {Bender} R.,  {Thomas} D.,  {Gebhardt} K.,
  {Magorrian} J.,  {Corsini} E.~M.,    {Wegner} G.,  2005, \mnras, 360, 1355

\bibitem[\protect\citeauthoryear{{Thomas}, {Saglia}, {Bender}, {Thomas},
  {Gebhardt}, {Magorrian}, {Corsini} \& {Wegner}}{{Thomas}
  et~al.}{2007}]{thomas+07}
{Thomas} J.,  {Saglia} R.~P.,  {Bender} R.,  {Thomas} D.,  {Gebhardt} K.,
  {Magorrian} J.,  {Corsini} E.~M.,    {Wegner} G.,  2007, \mnras, in press

\bibitem[\protect\citeauthoryear{{Tonry}, {Dressler}, {Blakeslee}, {Ajhar},
  {Fletcher}, {Luppino}, {Metzger} \& {Moore}}{{Tonry}
  et~al.}{2001}]{tonry_etal01}
{Tonry} J.~L.,  {Dressler} A.,  {Blakeslee} J.~P.,  {Ajhar} E.~A.,  {Fletcher}
  A.~B.,  {Luppino} G.~A.,  {Metzger} M.~R.,    {Moore} C.~B.,  2001, \apj,
  546, 681

\bibitem[\protect\citeauthoryear{{Tremblay}, {Merritt} \&
  {Williams}}{{Tremblay} et~al.}{1995}]{tremblay+95}
{Tremblay} B.,  {Merritt} D.,    {Williams} T.~B.,  1995, \apjl, 443, L5

\bibitem[\protect\citeauthoryear{{Treu} \& {Koopmans}}{{Treu} \&
  {Koopmans}}{2004}]{treu+koop04}
{Treu} T.,  {Koopmans} L.~V.~E.,  2004, \apj, 611, 739

\bibitem[\protect\citeauthoryear{{van der Marel} \& {Franx}}{{van der Marel} \&
  {Franx}}{1993}]{vdmarel_franx93}
{van der Marel} R.~P.,  {Franx} M.,  1993, \apj, 407, 525

\bibitem[\protect\citeauthoryear{{van der Marel}, {Rix}, {Carter}, {Franx},
  {White} \& {de Zeeuw}}{{van der Marel} et~al.}{1994}]{vdm1994}
{van der Marel} R.~P.,  {Rix} H.-W.,  {Carter} D.,  {Franx} M.,  {White}
  S.~D.~M.,    {de Zeeuw} P.~T.,  1994, \mnras, 268, 521

\end{thebibliography}

\appendix
\section{Photometric and kinematic data}
%
\begin{figure*}
\centering 
\includegraphics[width=0.90\hsize,angle=0.0]{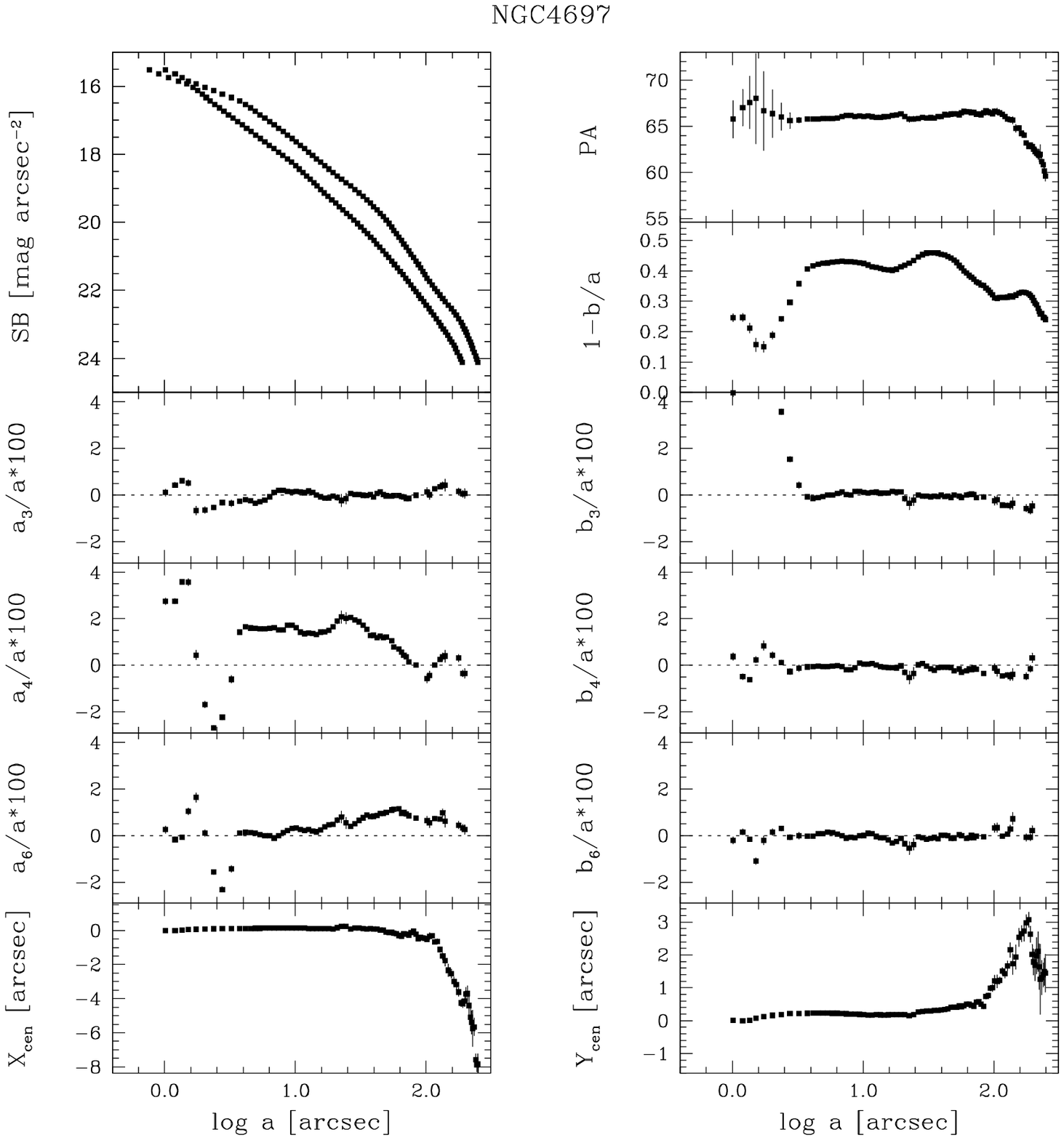}
\vskip0.5truecm
\caption[]{Isophotal parameters of NGC 4697 as a function of the
  logarithm of the semi-major axis distance in arcsec. The radial
  profiles of the R-band surface brightness, third, fourth, and sixth
  cosine Fourier coefficients ($a_3$, $a_4$, and $a_6$), and
  $x-$coordinate of the center $X_{cen}$ are plotted in the left
  panels (from top to bottom). The surface brightness is shown along
  the major (upper profile) and minor axis (lower profile). The radial
  profiles of the position angle (PA), ellipticity ($1-b/a$), third
  fourth, and sixth sine Fourier coefficients ($b_3$, $b_4$, and
  $b_6$), and $y-$coordinate of the center ($Y_{cen}$) are plotted in
  the right panels (from top to bottom).}
\label{fig:isophotes}
\end{figure*}

\newpage

\begin{sidewaystable*}
\begin{centering}
\label{tabphot4697}
\begin{tabular}{cccccccccccccc}
\multicolumn{14}{l}{Table \ref{tabphot4697}: Photometric parameters of NGC 4697}\\
\hline
   a                &   $\mu_R^a$         &         e        &       PA        &$\Delta x_c$ &$\Delta y_c$ & Err.$^b$ & $a_3/a$    & $b_3/a$      & $a_4/a$    & $b_4/a$    & $a_6/a$    & $b_6/a$     & Err.$^c$\\
$[$arcsec$]$            & $[$mag arcsec$^{-2}]$  &                  &      $[^\circ]$  & $[$arcsec$]$     & $[$arcsec$]$   & $[$arcsec$]$ & $\times100$ & $\times100$ & $\times100$ & $\times100$ & $\times100$ & $\times100$ & \\
\hline 
 1.013  $\pm$ 0.013 & 15.515$\pm$ 0.006 & 0.246$\pm$ 0.013 &  65.8$\pm$  2.0 & -0.012  & 0.007 &   0.009&   0.110 &   4.380 &   2.750 &   0.380 &    0.260 &  -0.210 &   0.162 \\
 1.201  $\pm$ 0.014 & 15.635$\pm$ 0.008 & 0.247$\pm$ 0.013 &  67.0$\pm$  2.0 &  0.000  & 0.000 &   0.010&   0.430 &   4.850 &   2.750 &  -0.490 &   -0.180 &   0.160 &   0.123 \\
 1.359  $\pm$ 0.020 & 15.753$\pm$ 0.007 & 0.212$\pm$ 0.017 &  67.6$\pm$  2.9 &  0.019  & 0.019 &   0.014&   0.620 &   6.790 &   3.580 &  -0.610 &   -0.080 &  -0.160 &   0.080 \\
 1.517  $\pm$ 0.029 & 15.847$\pm$ 0.008 & 0.157$\pm$ 0.023 &  68.0$\pm$  4.9 &  0.045  & 0.071 &   0.020&   0.520 &   9.620 &   3.570 &   0.240 &    1.030 &  -1.090 &   0.151 \\
 1.739  $\pm$ 0.028 & 15.929$\pm$ 0.008 & 0.151$\pm$ 0.019 &  66.7$\pm$  4.3 &  0.069  & 0.129 &   0.020&  -0.660 &   8.320 &   0.440 &   0.840 &    1.630 &  -0.200 &   0.206 \\
 2.023  $\pm$ 0.025 & 16.030$\pm$ 0.006 & 0.188$\pm$ 0.014 &  66.4$\pm$  2.6 &  0.083  & 0.162 &   0.018&  -0.650 &   6.000 &  -1.680 &   0.440 &    0.100 &   0.160 &   0.154 \\
 2.365  $\pm$ 0.021 & 16.134$\pm$ 0.005 & 0.242$\pm$ 0.009 &  66.0$\pm$  1.5 &  0.095  & 0.190 &   0.015&  -0.540 &   3.580 &  -2.690 &   0.130 &   -1.570 &   0.300 &   0.112 \\
 2.759  $\pm$ 0.017 & 16.233$\pm$ 0.004 & 0.297$\pm$ 0.006 &  65.6$\pm$  0.8 &  0.100  & 0.212 &   0.012&  -0.320 &   1.540 &  -2.220 &  -0.270 &   -2.310 &  -0.070 &   0.114 \\
 3.235  $\pm$ 0.009 & 16.333$\pm$ 0.003 & 0.358$\pm$ 0.003 &  65.7$\pm$  0.3 &  0.109  & 0.221 &   0.007&  -0.340 &   0.430 &  -0.600 &  -0.120 &   -1.430 &  -0.010 &   0.158 \\
 3.747  $\pm$ 0.007 & 16.436$\pm$ 0.002 & 0.406$\pm$ 0.001 &  65.8$\pm$  0.2 &  0.112  & 0.226 &   0.005&  -0.260 &  -0.070 &   1.420 &  -0.070 &    0.100 &  -0.030 &   0.023 \\
 4.138  $\pm$ 0.008 & 16.537$\pm$ 0.002 & 0.417$\pm$ 0.002 &  65.8$\pm$  0.2 &  0.114  & 0.228 &   0.006&  -0.200 &  -0.140 &   1.650 &  -0.060 &    0.140 &  -0.030 &   0.020 \\
 4.525  $\pm$ 0.008 & 16.637$\pm$ 0.002 & 0.422$\pm$ 0.001 &  65.8$\pm$  0.2 &  0.119  & 0.228 &   0.006&  -0.250 &  -0.110 &   1.600 &  -0.030 &    0.130 &   0.080 &   0.023 \\
 4.932  $\pm$ 0.009 & 16.737$\pm$ 0.002 & 0.425$\pm$ 0.001 &  65.8$\pm$  0.2 &  0.126  & 0.228 &   0.006&  -0.360 &  -0.060 &   1.590 &  -0.050 &    0.110 &   0.080 &   0.025 \\
 5.363  $\pm$ 0.009 & 16.837$\pm$ 0.002 & 0.426$\pm$ 0.001 &  65.9$\pm$  0.2 &  0.126  & 0.228 &   0.006&  -0.290 &  -0.000 &   1.560 &  -0.050 &    0.040 &   0.150 &   0.027 \\
 5.835  $\pm$ 0.009 & 16.937$\pm$ 0.002 & 0.428$\pm$ 0.001 &  65.8$\pm$  0.2 &  0.133  & 0.226 &   0.006&  -0.210 &   0.010 &   1.570 &  -0.030 &   -0.000 &   0.130 &   0.034 \\
 6.345  $\pm$ 0.010 & 17.037$\pm$ 0.002 & 0.430$\pm$ 0.001 &  65.9$\pm$  0.1 &  0.133  & 0.224 &   0.007&  -0.070 &   0.080 &   1.590 &  -0.020 &   -0.010 &   0.070 &   0.034 \\
 6.881  $\pm$ 0.011 & 17.137$\pm$ 0.002 & 0.431$\pm$ 0.001 &  66.1$\pm$  0.1 &  0.138  & 0.217 &   0.007&   0.120 &   0.130 &   1.620 &  -0.070 &   -0.110 &  -0.020 &   0.029 \\
 7.444  $\pm$ 0.011 & 17.237$\pm$ 0.001 & 0.431$\pm$ 0.001 &  66.2$\pm$  0.1 &  0.145  & 0.210 &   0.008&   0.210 &   0.040 &   1.520 &  -0.190 &   -0.010 &  -0.110 &   0.039 \\
 8.056  $\pm$ 0.012 & 17.337$\pm$ 0.001 & 0.430$\pm$ 0.001 &  66.2$\pm$  0.1 &  0.138  & 0.210 &   0.008&   0.210 &   0.000 &   1.530 &  -0.190 &    0.120 &  -0.070 &   0.039 \\
 8.706  $\pm$ 0.014 & 17.437$\pm$ 0.001 & 0.428$\pm$ 0.001 &  66.1$\pm$  0.2 &  0.138  & 0.198 &   0.010&   0.160 &   0.160 &   1.720 &  -0.110 &    0.210 &  -0.090 &   0.038 \\
 9.398  $\pm$ 0.016 & 17.536$\pm$ 0.001 & 0.427$\pm$ 0.001 &  66.1$\pm$  0.2 &  0.138  & 0.193 &   0.011&   0.150 &   0.150 &   1.720 &   0.100 &    0.300 &  -0.010 &   0.059 \\
10.136  $\pm$ 0.016 & 17.635$\pm$ 0.001 & 0.424$\pm$ 0.001 &  66.1$\pm$  0.2 &  0.140  & 0.186 &   0.011&   0.160 &   0.120 &   1.610 &   0.060 &    0.320 &   0.000 &   0.044 \\
10.900  $\pm$ 0.015 & 17.736$\pm$ 0.001 & 0.419$\pm$ 0.001 &  66.2$\pm$  0.1 &  0.138  & 0.176 &   0.011&   0.150 &   0.090 &   1.420 &   0.060 &    0.260 &   0.090 &   0.039 \\
11.706  $\pm$ 0.015 & 17.837$\pm$ 0.001 & 0.414$\pm$ 0.001 &  66.0$\pm$  0.1 &  0.131  & 0.174 &   0.011&   0.110 &   0.110 &   1.360 &   0.080 &    0.210 &   0.080 &   0.031 \\
12.598  $\pm$ 0.017 & 17.937$\pm$ 0.001 & 0.411$\pm$ 0.001 &  66.0$\pm$  0.1 &  0.112  & 0.181 &   0.012&   0.170 &   0.130 &   1.400 &   0.020 &    0.250 &  -0.040 &   0.052 \\
13.531  $\pm$ 0.018 & 18.036$\pm$ 0.001 & 0.407$\pm$ 0.001 &  66.0$\pm$  0.1 &  0.117  & 0.176 &   0.013&   0.110 &   0.080 &   1.360 &  -0.060 &    0.200 &  -0.090 &   0.039 \\
14.530  $\pm$ 0.018 & 18.136$\pm$ 0.001 & 0.405$\pm$ 0.001 &  66.0$\pm$  0.1 &  0.119  & 0.174 &   0.013&   0.010 &   0.090 &   1.320 &  -0.070 &    0.160 &  -0.100 &   0.028 \\
15.595  $\pm$ 0.022 & 18.236$\pm$ 0.001 & 0.403$\pm$ 0.001 &  66.1$\pm$  0.1 &  0.121  & 0.178 &   0.016&  -0.070 &   0.100 &   1.420 &  -0.090 &    0.250 &  -0.190 &   0.030 \\
16.761  $\pm$ 0.027 & 18.335$\pm$ 0.001 & 0.402$\pm$ 0.001 &  66.2$\pm$  0.2 &  0.107  & 0.188 &   0.019&  -0.130 &   0.160 &   1.430 &  -0.110 &    0.370 &  -0.320 &   0.050 \\
18.038  $\pm$ 0.029 & 18.435$\pm$ 0.001 & 0.405$\pm$ 0.001 &  66.2$\pm$  0.2 &  0.105  & 0.193 &   0.021&  -0.130 &   0.140 &   1.500 &  -0.110 &    0.470 &  -0.260 &   0.039 \\
19.477  $\pm$ 0.034 & 18.537$\pm$ 0.001 & 0.413$\pm$ 0.001 &  66.4$\pm$  0.2 &  0.088  & 0.193 &   0.024&  -0.050 &   0.140 &   1.660 &  -0.020 &    0.490 &  -0.120 &   0.034 \\
20.955  $\pm$ 0.049 & 18.637$\pm$ 0.001 & 0.419$\pm$ 0.002 &  66.1$\pm$  0.2 &  0.157  & 0.181 &   0.035&  -0.100 &  -0.140 &   1.890 &  -0.300 &    0.670 &  -0.350 &   0.130 \\
22.554  $\pm$ 0.074 & 18.735$\pm$ 0.001 & 0.425$\pm$ 0.003 &  65.8$\pm$  0.3 &  0.248  & 0.162 &   0.053&  -0.230 &  -0.350 &   2.070 &  -0.520 &    0.790 &  -0.540 &   0.270 \\
24.352  $\pm$ 0.078 & 18.835$\pm$ 0.001 & 0.433$\pm$ 0.003 &  65.8$\pm$  0.3 &  0.228  & 0.186 &   0.055&  -0.150 &  -0.220 &   2.020 &  -0.340 &    0.540 &  -0.380 &   0.254 \\
26.452  $\pm$ 0.056 & 18.936$\pm$ 0.001 & 0.446$\pm$ 0.002 &  65.9$\pm$  0.2 &  0.093  & 0.264 &   0.040&   0.070 &  -0.000 &   2.070 &   0.020 &    0.400 &  -0.050 &   0.070 \\
28.543  $\pm$ 0.057 & 19.036$\pm$ 0.001 & 0.453$\pm$ 0.001 &  66.0$\pm$  0.2 &  0.150  & 0.274 &   0.040&   0.030 &  -0.040 &   1.940 &   0.080 &    0.510 &  -0.090 &   0.077 \\
30.687  $\pm$ 0.060 & 19.135$\pm$ 0.001 & 0.458$\pm$ 0.001 &  65.9$\pm$  0.2 &  0.147  & 0.290 &   0.042&   0.020 &  -0.050 &   1.870 &  -0.040 &    0.650 &  -0.150 &   0.059 \\
32.840  $\pm$ 0.060 & 19.236$\pm$ 0.001 & 0.459$\pm$ 0.001 &  65.9$\pm$  0.2 &  0.124  & 0.302 &   0.042&  -0.010 &  -0.070 &   1.720 &  -0.140 &    0.760 &  -0.120 &   0.044 \\
35.061  $\pm$ 0.062 & 19.336$\pm$ 0.001 & 0.459$\pm$ 0.001 &  65.9$\pm$  0.2 &  0.088  & 0.312 &   0.044&  -0.020 &  -0.040 &   1.550 &  -0.210 &    0.880 &  -0.090 &   0.045 \\
37.385  $\pm$ 0.059 & 19.435$\pm$ 0.001 & 0.458$\pm$ 0.001 &  66.1$\pm$  0.1 &  0.102  & 0.314 &   0.042&   0.030 &  -0.060 &   1.280 &  -0.100 &    0.810 &  -0.140 &   0.066 \\
39.678  $\pm$ 0.060 & 19.536$\pm$ 0.001 & 0.455$\pm$ 0.001 &  66.1$\pm$  0.1 &  0.078  & 0.326 &   0.042&  -0.080 &   0.010 &   1.290 &  -0.080 &    0.830 &   0.020 &   0.050 \\
42.182  $\pm$ 0.066 & 19.635$\pm$ 0.001 & 0.453$\pm$ 0.001 &  66.3$\pm$  0.1 &  0.074  & 0.343 &   0.046&   0.070 &  -0.060 &   1.190 &  -0.070 &    0.860 &   0.020 &   0.071 \\
44.663  $\pm$ 0.072 & 19.734$\pm$ 0.001 & 0.448$\pm$ 0.001 &  66.3$\pm$  0.2 &  0.028  & 0.359 &   0.051&   0.140 &  -0.040 &   1.250 &  -0.070 &    0.940 &  -0.000 &   0.062 \\
47.108  $\pm$ 0.073 & 19.834$\pm$ 0.001 & 0.442$\pm$ 0.001 &  66.3$\pm$  0.1 &  0.010  & 0.402 &   0.052&   0.020 &  -0.100 &   1.190 &  -0.100 &    0.950 &  -0.120 &   0.074 \\
49.534  $\pm$ 0.078 & 19.934$\pm$ 0.001 & 0.436$\pm$ 0.001 &  66.3$\pm$  0.2 & -0.088  & 0.409 &   0.055&  -0.030 &   0.010 &   1.210 &  -0.240 &    1.000 &  -0.110 &   0.045 \\
\hline
\end{tabular}
\end{centering}
\end{sidewaystable*}

\begin{sidewaystable*}
\begin{centering}
\label{tabphot4697a}
\begin{tabular}{cccccccccccccc}
\hline
   a                &   $\mu_R^a$         &         e        &       PA        &$\Delta x_c$ &$\Delta y_c$ & Err.$^b$ & $a_3/a$    & $b_3/a$      & $a_4/a$    & $b_4/a$    & $a_6/a$    & $b_6/a$     & Err.$^c$\\
$[$arcsec$]$            & $[$mag arcsec$^{-2}]$  &                  &      $[^\circ]$  & $[$arcsec$]$     & $[$arcsec$]$   & $[$arcsec$]$ & $\times100$ & $\times100$ & $\times100$ & $\times100$ & $\times100$ & $\times100$ & \\
\hline 
51.950  $\pm$ 0.083 & 20.036$\pm$ 0.001 & 0.428$\pm$ 0.001 &  66.4$\pm$  0.2 & -0.119  & 0.402 &   0.059&    ...  &    ...  &    ...  &    ...  &     ...  &    ...  &   0.045 \\
54.308  $\pm$ 0.087 & 20.135$\pm$ 0.001 & 0.417$\pm$ 0.001 &  66.2$\pm$  0.2 & -0.098  & 0.459 &   0.062&  -0.040 &  -0.040 &   1.060 &  -0.160 &    1.090 &   0.030 &   0.062 \\
56.753  $\pm$ 0.092 & 20.235$\pm$ 0.001 & 0.409$\pm$ 0.001 &  66.5$\pm$  0.2 & -0.167  & 0.424 &   0.065&  -0.020 &  -0.110 &   0.790 &  -0.290 &    1.130 &  -0.020 &   0.065 \\
59.253  $\pm$ 0.092 & 20.336$\pm$ 0.001 & 0.399$\pm$ 0.001 &  66.6$\pm$  0.2 & -0.198  & 0.452 &   0.065&    ...  &    ...  &    ...  &    ...  &     ...  &    ...  &   0.065 \\
62.028  $\pm$ 0.092 & 20.436$\pm$ 0.001 & 0.393$\pm$ 0.001 &  66.6$\pm$  0.2 & -0.298  & 0.502 &   0.065&  -0.010 &   0.000 &   0.700 &  -0.240 &    1.150 &  -0.130 &   0.059 \\
64.741  $\pm$ 0.082 & 20.536$\pm$ 0.001 & 0.385$\pm$ 0.001 &  66.6$\pm$  0.1 & -0.345  & 0.505 &   0.058&  -0.050 &   0.040 &   0.570 &  -0.160 &    0.960 &  -0.100 &   0.068 \\
67.520  $\pm$ 0.078 & 20.635$\pm$ 0.001 & 0.379$\pm$ 0.001 &  66.5$\pm$  0.1 & -0.205  & 0.476 &   0.055&  -0.090 &   0.050 &   0.440 &  -0.190 &    0.990 &   0.010 &   0.048 \\
70.445  $\pm$ 0.085 & 20.734$\pm$ 0.001 & 0.372$\pm$ 0.001 &  66.5$\pm$  0.1 & -0.219  & 0.438 &   0.060&  -0.140 &   0.030 &   0.360 &  -0.110 &    0.920 &  -0.090 &   0.064 \\
73.422  $\pm$ 0.090 & 20.834$\pm$ 0.001 & 0.364$\pm$ 0.001 &  66.3$\pm$  0.1 & -0.279  & 0.567 &   0.064&  -0.140 &  -0.110 &   0.140 &  -0.170 &    0.850 &  -0.040 &   0.078 \\
76.678  $\pm$ 0.085 & 20.936$\pm$ 0.001 & 0.360$\pm$ 0.001 &  66.4$\pm$  0.1 & -0.140  & 0.583 &   0.060&    ...  &    ...  &    ...  &    ...  &     ...  &    ...  &   0.078 \\
80.086  $\pm$ 0.093 & 21.037$\pm$ 0.001 & 0.354$\pm$ 0.001 &  66.3$\pm$  0.1 & -0.050  & 0.497 &   0.066&    ...  &    ...  &    ...  &    ...  &     ...  &    ...  &   0.078 \\
83.687  $\pm$ 0.109 & 21.138$\pm$ 0.001 & 0.351$\pm$ 0.001 &  66.4$\pm$  0.2 & -0.228  & 0.436 &   0.077&  -0.010 &  -0.080 &   0.010 &  -0.350 &    0.750 &  -0.060 &   0.094 \\
87.231  $\pm$ 0.101 & 21.239$\pm$ 0.001 & 0.344$\pm$ 0.001 &  66.7$\pm$  0.1 & -0.502  & 0.738 &   0.071&    ...  &    ...  &    ...  &    ...  &     ...  &    ...  &   0.094 \\
90.680  $\pm$ 0.120 & 21.339$\pm$ 0.001 & 0.336$\pm$ 0.001 &  66.6$\pm$  0.2 & -0.398  & 0.769 &   0.085&    ...  &    ...  &    ...  &    ...  &     ...  &    ...  &   0.094 \\
94.336  $\pm$ 0.135 & 21.439$\pm$ 0.001 & 0.329$\pm$ 0.001 &  66.4$\pm$  0.2 & -0.445  & 0.992 &   0.096&    ...  &    ...  &    ...  &    ...  &     ...  &    ...  &   0.094 \\
98.111  $\pm$ 0.165 & 21.538$\pm$ 0.001 & 0.321$\pm$ 0.002 &  66.3$\pm$  0.2 & -0.457  & 1.004 &   0.117&    ...  &    ...  &    ...  &    ...  &     ...  &    ...  &   0.094 \\
101.661 $\pm$ 0.232 & 21.637$\pm$ 0.001 & 0.311$\pm$ 0.002 &  66.6$\pm$  0.3 & -0.509  & 1.214 &   0.164&   0.140 &  -0.240 &  -0.560 &  -0.110 &    0.640 &   0.340 &   0.200 \\
105.981 $\pm$ 0.288 & 21.737$\pm$ 0.001 & 0.309$\pm$ 0.003 &  66.6$\pm$  0.4 & -0.347  & 1.197 &   0.203&   0.020 &  -0.200 &  -0.440 &  -0.260 &    0.540 &   0.340 &   0.210 \\
111.325 $\pm$ 0.162 & 21.841$\pm$ 0.001 & 0.312$\pm$ 0.001 &  66.4$\pm$  0.2 & -0.319  & 1.230 &   0.115&    ...  &    ...  &    ...  &    ...  &     ...  &    ...  &   0.210 \\
116.502 $\pm$ 0.166 & 21.944$\pm$ 0.001 & 0.313$\pm$ 0.001 &  66.3$\pm$  0.2 & -0.664  & 1.511 &   0.117&   0.270 &  -0.430 &   0.000 &  -0.450 &    0.740 &  -0.040 &   0.092 \\
121.793 $\pm$ 0.201 & 22.043$\pm$ 0.001 & 0.313$\pm$ 0.002 &  66.1$\pm$  0.2 & -0.652  & 1.433 &   0.142&    ...  &    ...  &    ...  &    ...  &     ...  &    ...  &   0.092 \\
127.455 $\pm$ 0.218 & 22.138$\pm$ 0.001 & 0.313$\pm$ 0.002 &  65.8$\pm$  0.2 & -1.119  & 1.668 &   0.154&   0.370 &  -0.440 &   0.260 &  -0.410 &    0.710 &   0.070 &   0.099 \\
133.917 $\pm$ 0.324 & 22.234$\pm$ 0.001 & 0.318$\pm$ 0.002 &  65.7$\pm$  0.3 & -1.509  & 2.154 &   0.229&   0.420 &  -0.440 &   0.370 &  -0.480 &    0.970 &   0.290 &   0.170 \\
139.843 $\pm$ 0.460 & 22.331$\pm$ 0.001 & 0.314$\pm$ 0.003 &  65.7$\pm$  0.4 & -1.775  & 1.725 &   0.325&   0.430 &  -0.340 &   0.400 &  -0.400 &    0.620 &   0.720 &   0.265 \\
147.198 $\pm$ 0.536 & 22.430$\pm$ 0.001 & 0.320$\pm$ 0.003 &  64.8$\pm$  0.5 & -2.335  & 1.937 &   0.379&    ...  &    ...  &    ...  &    ...  &     ...  &    ...  &   0.265 \\
155.523 $\pm$ 0.400 & 22.533$\pm$ 0.001 & 0.326$\pm$ 0.002 &  64.8$\pm$  0.3 & -2.535  & 2.537 &   0.283&    ...  &    ...  &    ...  &    ...  &     ...  &    ...  &   0.265 \\
163.016 $\pm$ 0.359 & 22.633$\pm$ 0.001 & 0.329$\pm$ 0.002 &  64.2$\pm$  0.3 & -2.982  & 2.651 &   0.254&    ...  &    ...  &    ...  &    ...  &     ...  &    ...  &   0.265 \\
170.075 $\pm$ 0.429 & 22.728$\pm$ 0.001 & 0.329$\pm$ 0.002 &  64.1$\pm$  0.3 & -3.165  & 2.720 &   0.304&    ...  &    ...  &    ...  &    ...  &     ...  &    ...  &   0.265 \\
177.623 $\pm$ 0.365 & 22.827$\pm$ 0.001 & 0.328$\pm$ 0.002 &  63.2$\pm$  0.2 & -3.611  & 2.977 &   0.258&   0.160 &  -0.570 &   0.320 &  -0.470 &    0.440 &  -0.080 &   0.163 \\
185.102 $\pm$ 0.334 & 22.934$\pm$ 0.001 & 0.325$\pm$ 0.002 &  62.8$\pm$  0.2 & -4.251  & 3.075 &   0.237&    ...  &    ...  &    ...  &    ...  &     ...  &    ...  &   0.163 \\
190.987 $\pm$ 0.358 & 23.036$\pm$ 0.001 & 0.320$\pm$ 0.002 &  63.0$\pm$  0.2 & -4.284  & 2.625 &   0.253&   0.040 &  -0.670 &  -0.350 &  -0.140 &    0.320 &  -0.070 &   0.153 \\
196.718 $\pm$ 0.432 & 23.132$\pm$ 0.001 & 0.312$\pm$ 0.002 &  62.8$\pm$  0.3 & -4.146  & 2.013 &   0.305&   0.070 &  -0.470 &  -0.350 &   0.330 &    0.260 &   0.220 &   0.200 \\
202.089 $\pm$ 0.559 & 23.227$\pm$ 0.001 & 0.303$\pm$ 0.003 &  62.5$\pm$  0.4 & -3.715  & 1.787 &   0.395&    ...  &    ...  &    ...  &    ...  &     ...  &    ...  &   0.200 \\
207.354 $\pm$ 0.675 & 23.323$\pm$ 0.001 & 0.294$\pm$ 0.003 &  62.2$\pm$  0.5 & -3.703  & 1.692 &   0.478&    ...  &    ...  &    ...  &    ...  &     ...  &    ...  &   0.200 \\
212.802 $\pm$ 0.787 & 23.421$\pm$ 0.001 & 0.290$\pm$ 0.004 &  62.1$\pm$  0.5 & -4.401  & 1.973 &   0.556&    ...  &    ...  &    ...  &    ...  &     ...  &    ...  &   0.200 \\
217.940 $\pm$ 0.851 & 23.521$\pm$ 0.001 & 0.277$\pm$ 0.004 &  62.0$\pm$  0.6 & -5.084  & 2.106 &   0.602&    ...  &    ...  &    ...  &    ...  &     ...  &    ...  &   0.200 \\
222.191 $\pm$ 0.996 & 23.613$\pm$ 0.001 & 0.266$\pm$ 0.005 &  61.8$\pm$  0.7 & -5.386  & 1.642 &   0.704&    ...  &    ...  &    ...  &    ...  &     ...  &    ...  &   0.200 \\
226.618 $\pm$ 1.524 & 23.707$\pm$ 0.001 & 0.258$\pm$ 0.007 &  62.0$\pm$  1.1 & -5.745  & 1.259 &   1.078&    ...  &    ...  &    ...  &    ...  &     ...  &    ...  &   0.200 \\
233.406 $\pm$ 0.749 & 23.815$\pm$ 0.001 & 0.260$\pm$ 0.003 &  61.2$\pm$  0.5 & -5.653  & 1.309 &   0.529&    ...  &    ...  &    ...  &    ...  &     ...  &    ...  &   0.200 \\
238.734 $\pm$ 0.519 & 23.925$\pm$ 0.001 & 0.247$\pm$ 0.002 &  60.9$\pm$  0.4 & -7.590  & 1.397 &   0.367&    ...  &    ...  &    ...  &    ...  &     ...  &    ...  &   0.200 \\
243.821 $\pm$ 0.590 & 24.020$\pm$ 0.001 & 0.246$\pm$ 0.003 &  60.2$\pm$  0.4 & -7.878  & 1.488 &   0.417&    ...  &    ...  &    ...  &    ...  &     ...  &    ...  &   0.200 \\
247.975 $\pm$ 0.827 & 24.104$\pm$ 0.001 & 0.240$\pm$ 0.004 &  59.6$\pm$  0.6 & -7.811  & 1.447 &   0.585&    ...  &    ...  &    ...  &    ...  &     ...  &    ...  &   0.200 \\
253.321 $\pm$ 0.972 & 24.189$\pm$ 0.001 & 0.245$\pm$ 0.004 &  59.4$\pm$  0.6 & -7.540  & 1.035 &   0.687&    ...  &    ...  &    ...  &    ...  &     ...  &    ...  &   0.200 \\
\hline
\multicolumn{14}{l}{$^a$ Statistical errors not including systematics due to photometric calibration and sky subtraction}. \\
\multicolumn{14}{l}{$^b$ Error on the center coordinates from the residual rms of the ellipse fit to the isophotes:
Err=rms(fit)$/\sqrt{N}$ with $N\le128$ the number of fitted points of the isophotes.}\\
\multicolumn{14}{l}{$^c$ Error of Fourier coefficients defined as Err$=\sqrt{\frac{\sum_{i=10}^{N/2}(a_i^2+b_i^2)}{N/2-10}}\times\frac{100}{a}$.}
\end{tabular}
\end{centering}
\end{sidewaystable*}

\begin{centering}
\begin{table*}
\caption{The kinematics of NGC 4697 along the major axis (P.A.=66$^\circ$).
 Positive radii are to the north-east.}
\label{tabn4697mj}
\begin{tabular}{rrrrrrrrr}
\hline
R & V~  & dV & $\sigma$~ & $d\sigma$~ & $h_3$~ & $dh_3$~ & $h_4$~ & $dh_4$~ \\
$(")$ & (km/s)   & (km/s)   & (km/s) & (km/s) & & & & \\
\hline  
       0.38   &    -13.2   &     0.4  &       180.0   &     0.4  &   -0.004   &    0.002   &    0.007   &    0.002 \\
       0.97   &    -39.5   &     0.2  &     175.3   &     0.4  &    0.027   &    0.002   &    0.014   &    0.002 \\
       1.57   &     -51.8   &     0.2  &     174.4   &      0.4  &    0.041   &    0.002   &    0.017   &    0.002 \\
       2.27   &    -70.1   &     0.3  &     169.4   &     0.4  &    0.049   &    0.002   &    0.029   &    0.002 \\
       3.16   &    -83.1   &      0.3  &     163.4   &     0.4  &     0.08   &    0.002   &    0.037   &    0.002 \\
        4.26   &    -89.9   &     0.3  &     160.6   &     0.4  &    0.095   &    0.002   &    0.034   &    0.002 \\
       5.55   &     -93.8   &     0.4  &     157.9   &     0.5  &    0.105   &    0.003   &    0.037   &    0.003 \\
       7.14   &    -97.3   &     0.4  &     162.3   &     0.5  &     0.12   &    0.002   &    0.036   &    0.002 \\
       9.22   &    -94.2   &     0.4  &     163.3   &     0.6  &    0.096   &    0.002   &    0.019   &    0.003 \\
       12.00   &    -96.6   &     0.4  &     166.3   &     0.5  &    0.093   &    0.002   &     0.010   &    0.002 \\
       15.67   &    -95.5   &     0.4  &     170.5   &     0.6  &     0.080   &    0.002   &    0.004   &    0.003 \\
       20.63   &    -114.1   &     0.6  &     160.3   &     0.8  &     0.150   &    0.003   &    0.022   &    0.004 \\
       27.62   &    -109.1   &     0.6  &     157.8   &     0.8  &     0.140   &    0.003   &   -0.016   &    0.004 \\
       38.07   &    -115.1   &     0.7  &     151.9   &     1.0  &    0.122   &    0.004   &   -0.003   &    0.005 \\
       58.05   &    -108.6   &     1.0  &     143.2   &     1.4  &    0.122   &    0.006   &   -0.029   &    0.007 \\
       92.52   &    -111.5   &     2.9  &     140.9   &     3.7  &   -0.017   &     0.020   &   -0.082   &    0.015 \\
       -0.22  &     7.0   &     0.4  &     179.6   &     0.4  &    -0.024   &    0.002   &    0.015   &    0.002 \\
       -0.81  &     31.3   &     0.3  &       177.0   &     0.4  &    -0.042   &    0.002   &    0.026   &    0.001 \\
       -1.41  &     51.9   &     0.3  &     173.3   &     0.4  &    -0.061   &    0.002   &    0.029   &    0.001 \\
       -2.11  &      70.1   &     0.3  &       169.0   &     0.5  &    -0.072   &    0.002   &    0.045   &    0.002 \\
       -3.00  &     85.6   &     0.3  &     163.4   &     0.5  &      -0.100   &    0.002   &    0.052   &    0.002 \\
       -4.00  &     92.8   &      0.3  &     161.8   &     0.3  &    -0.112   &    0.001   &    0.034   &    0.002 \\
        -5.19  &     93.6   &      0.3  &     162.7   &     0.3  &    -0.112   &    0.001   &    0.041   &    0.002 \\
       -6.69  &     98.0   &     0.3  &     160.3   &     0.3  &    -0.135   &    0.001   &    0.043   &    0.002 \\
       -8.57  &     96.0   &      0.3  &     159.1   &     0.3  &    -0.114   &    0.001   &    0.034   &    0.002 \\
       -11.05  &     98.3   &     0.3  &     165.8   &     0.3  &    -0.131   &    0.001   &     0.030   &    0.002 \\
       -14.32  &     93.2   &     0.3  &     172.2   &     0.3  &    -0.126   &    0.001   &   -0.014   &    0.002 \\
       -18.79  &     103.9   &     0.3  &     172.1   &     0.3  &    -0.145   &    0.001   &   -0.011   &    0.002 \\
       -24.99  &     112.6   &     0.4  &     168.1   &     0.3  &     -0.170   &    0.002   &    0.005   &    0.002 \\
          -34  &     120.1   &     0.5  &       160.0   &     0.5  &    -0.168   &    0.002   &   -0.026   &    0.003 \\
       -48.87  &     113.7   &     0.7  &     152.5   &     0.6  &    -0.132   &    0.004   &   -0.018   &    0.004 \\
       -76.65  &     115.2   &     1.9  &     153.1   &      2.1  &    -0.087   &    0.012   &   -0.004   &     0.010 \\
\hline
\end{tabular}
\end{table*}
\end{centering}

\begin{centering}
\begin{table*}
\caption{The kinematics of NGC 4697 along the minor axis (P.A.=156$^\circ$).
 Positive radii are to the south-east.}
\label{tabn4697mn}
\begin{tabular}{rrrrrrrrr}
\hline
R & V~  & dV & $\sigma$~ & $d\sigma$~ & $h_3$~ & $dh_3$~ & $h_4$~ & $dh_4$~ \\
$(")$ & (km/s)   & (km/s)   & (km/s) & (km/s) & & & & \\
\hline
       0.09  &    -1.3  &      0.2  &     186.3   &     0.2  &    -0.014  &     0.001  &     0.011   &    0.001 \\
       0.49  &     1.6  &      0.3  &     183.7   &     0.3  &    -0.001  &     0.001  &     0.006   &    0.001 \\
        1.00  &     2.3  &      0.1  &     179.8   &     0.1  &    -0.013  &     0.005  &     0.009   &    0.001 \\
       1.58  &    0.3  &      0.3  &     176.4   &     0.3  &    -0.007  &     0.001  &     0.005   &    0.001 \\
       2.27  &     1.6  &       0.3  &     177.9   &     0.3  &     0.001  &     0.001  &     0.037   &    0.001 \\
       3.25  &    -1.3  &      0.2  &     180.4   &     0.2  &    -0.001  &     0.001  &      0.030   &    0.001 \\
       4.73  &     4.1  &      0.3  &     184.3   &      0.3  &    -0.006  &     0.001  &     0.034   &    0.001 \\
       6.98  &     1.9  &      0.3  &     178.1   &     0.3  &    -0.019  &     0.001  &     0.015   &    0.001 \\
       10.66  &    -4.2  &      0.5  &     177.6   &     0.5  &    -0.022  &     0.002  &     0.023   &    0.002 \\
       17.37  &     5.2  &      0.7  &     175.9   &     0.7  &    -0.007  &     0.003  &     0.005   &    0.002 \\
       31.77  &    -4.6  &      1.4  &     173.2   &     1.4  &    -0.005  &     0.006  &    -0.012   &    0.005 \\
      -0.31  &    0.6  &      0.2  &       185.0   &     0.2  &      0.020 &     0.001  &     0.004   &    0.001 \\
      -0.80  &    1.0  &      0.1  &     181.9   &     0.1  &     0.018  &     0.005  &     0.006   &    0.001 \\
      -1.40  &    -1.4  &      0.2  &     176.5   &     0.2  &     0.018  &     0.001  &     0.021   &    0.001 \\
      -2.18  &   -0.8  &      0.3  &     178.9   &     0.3  &     0.013  &     0.001  &     0.027   &    0.001 \\
      -3.36  &    -1.9  &      0.3  &     176.7   &     0.3  &    -0.004  &     0.001  &     0.023   &    0.001 \\
      -5.13  &   -0.8  &      0.3  &       175.0   &      0.3  &      0.020  &     0.001  &     0.018   &    0.001 \\
      -7.95  &     1.0  &      0.4  &     175.2   &     0.5  &      0.020  &     0.002  &      0.050   &    0.002 \\
      -12.74  &    0.6  &      0.6  &     173.2   &     0.7  &     0.015  &     0.003  &     0.037   &    0.003 \\
      -22.55  &     1.9  &      0.9  &     169.7   &     1.0  &     0.017  &     0.004  &    -0.001   &    0.003 \\
      -49.42  &     4.4  &      2.1  &     158.1   &     2.3  &     0.001  &      0.010  &    -0.011   &    0.007 \\
\hline
\end{tabular}
\end{table*}
\end{centering}

\label{lastpage} 

\end{document}